\documentclass[review]{elsarticle}

\usepackage{hyperref}
\usepackage{graphics}
\usepackage{amsmath}
\usepackage{amssymb}
\usepackage{verbatim}

\journal{Astroparticle Physics}
\begin{document}
\begin{frontmatter}

\title{ Soft Particle Production in Very High Energy Hadron Interactions }

\author{Jan Ebr\corref{cor1}}
\ead{ebr@fzu.cz}
\cortext[cor1]{Corresponding author}
\author{Petr Ne\v{c}esal\corref{}}
\author{Jan Ridky\corref{}}
\address{Institute of Physics of the Czech Academy of Sciences, Na Slovance 1999/2, 18221 Prague 8, Czech
Republic}
\begin{abstract}
Indications of a discrepancy between simulations and data on the number of muons in cosmic ray (CR) showers exist over a large span of energies. We focus in particular on the excess of multi-muon bundles observed by the DELPHI detector at LEP and on the excess in the muon number in general reported by the Pierre Auger Observatory. Even though the primary CR energies relevant for these experiments differ by orders of magnitude, we can find a single mechanism which can simultaneously increase predicted muon counts for both, while not violating constraints from accelerators or from the longitudinal shower development as observed by the Pierre Auger Observatory. We present a brief motivation and describe a practical implementation of such a model, based on the addition of soft particles to interactions above a chosen energy threshold. Results of an extensive set of simulations show the behavior of this model in various parts of a simplified parameter space.
\end{abstract}


\end{frontmatter}

\section{Motivation}

The cosmic ray (CR) showers originated by cosmic particles with energies up to the limit of the order of $10^{20}$ eV provide a unique opportunity to study particle interactions at energies inaccessible to terrestrial accelerators. Dedicated models of hadronic interactions (e.g. \cite{QGSII-4,EPOS-LHC,Sibyll}) describe features of these showers remarkably well with the exception of muon production. Discrepancies between the data and the models have already been observed in CR showers detected by three out of the four LEP experiments which in addition to $e^{+}e^{-}$ interactions, measured also cosmic muons, each of them with different overburden. The shallowest experiment L3+C \cite{L3C2} with 30 m of overburden used simultaneously a shower array on the ground. Detectors designed for collider experiments provided data superior to standard cosmic ray experiments and their tracking capabilities combined with the overburden enabled measurements of the muon content of the cores of showers at energies $10^{14}-10^{17}$ eV. ALEPH \cite{ALEPH} and L3+C were able to measure individual muon tracks up to saturation while DELPHI \cite{DELPHI} made use of the fine granularity of its hadron calorimeter and measured integrated muon multiplicities. All three experiments observed high multiplicity events with a frequency which could not be fully described even by pure iron composition. The DELPHI results are the most robust from the statistical point of view.

The discrepancy between these measurements of high-multiplicity muon bundles and the models have inspired other experiments to investigate the issue. ALICE \cite{ALICE2} has conducted dedicated cosmic-ray measurements similar do DELPHI and found results consistent with current hadronic interaction models.  ALICE has better resolution in muon multiplicity then DELPHI, but smaller detection area and a smaller cut-off – only 16 GeV for vertical muons, compared to 52 GeV for DELPHI; thus those results are not directly comparable. Another experiment dedicated to muon bundles is NEVOD-DECOR\cite{NEVOD} which however does not measure individual high-energy muons directly.

The influx of new data from LHC experiments allowed the tuning of models up to these energies. However, the tuned CR models still do not describe well, in particular, the DELPHI data as discussed later. Generally the models tuned to LHC do better than their earlier versions - e.g. QGSJET-II-03 \cite{QGSII-3} vs. QGSJET-II-04 \cite{QGSII-4}. However, they also tend to underestimate the muon component of ultra-high energy CR showers as indicated chiefly by several studies done at the Pierre Auger Observatory \cite{AUGER}, which can access very high c.m.s. energies thanks to its sensitivity to ultra-high energy CR \cite{AUGERmu1,AUGERmu2,AUGERmu3}. The available detection techniques of CR showers allow us to study only the gross features of the most common interactions. The ubiquity of the muon excess and span of the energy range where it occurs suggest that its origin is more probably connected with standard features of hadronic interactions rather then with some new exotic phenomena. 

This paper aims to show that the addition of particles (mainly pions) with small momenta in the corresponding c.m. system, ("soft particles" from now on) to high-energy interactions in the CR showers could improve the description of the muon production in CR showers without significantly deviating from the framework of standard modeling of hadron interactions. Interestingly enough it turns out that if such an effect is assumed, the constraints provided by LHC experiments, the DELPHI CR measurements and the data of the Pierre Auger Observatory are sufficiently strong and do not leave much space for significant changes of the models because of the large energy span where the muon excess is consistently observed.

Obviously, the addition of soft particles is not the only possible explanation for the observed excess. Barring exotic physics, the main other possible source of muons is the uncertainty in the production of the heavy flavors \cite{Nosek1}. However, it is now experimentally established by LHC that the heavy flavor production does not suffice to increase the muon content in the showers detected by DELPHI \cite{Charm} to observable amount. Even at very high energies where one could hypothesize copious heavy flavor production this would actually lead to much larger missing energy and lower muon production as the hard muons and neutrinos produced via heavy flavors would take away energy missing in soft pion production \cite{Necesal}. 

\section{DELPHI Data and Simulations}\label{delphi}

The main cosmic ray result from DELPHI is the measurement of bundles of muons from extensive air showers. At high observed muon multiplicities, the observed flux of events is in excess with respect to simulations even for a pure iron primary beam. Actually, due to saturation of the signal in streamer tubes -- the smallest sensing units in the hadron calorimeter of DELPHI (HCAL), which signaled just passing one or more muons, DELPHI measured a lower limit of muon multiplicity in each event. Only at low multiplicities this coincides with the real muon multiplicity in HCAL. The standard detector simulation program DELSIM \cite{DELSIM} traces all particles produced in $e^+e^-$ interactions through the whole volume of the DELPHI detector. This would lead to prohibitive CPU times given the amount of simulated showers. We have thus created a simplified DELPHI model of muon detection which we use to interpret results of CORSIKA simulations in order to be able to compare the DELPHI data first with different new hadronic interaction models and ultimately with simulations having added soft particles. As this approach ignores many details of the experiment, it would be extremely difficult to reliably predict the overall normalization of the flux of the muon bundles and to understand which effects are due to the change of hadronic model and which simply due to the inadequacy of our simulations. 

\begin{figure}

\includegraphics[width=9cm]{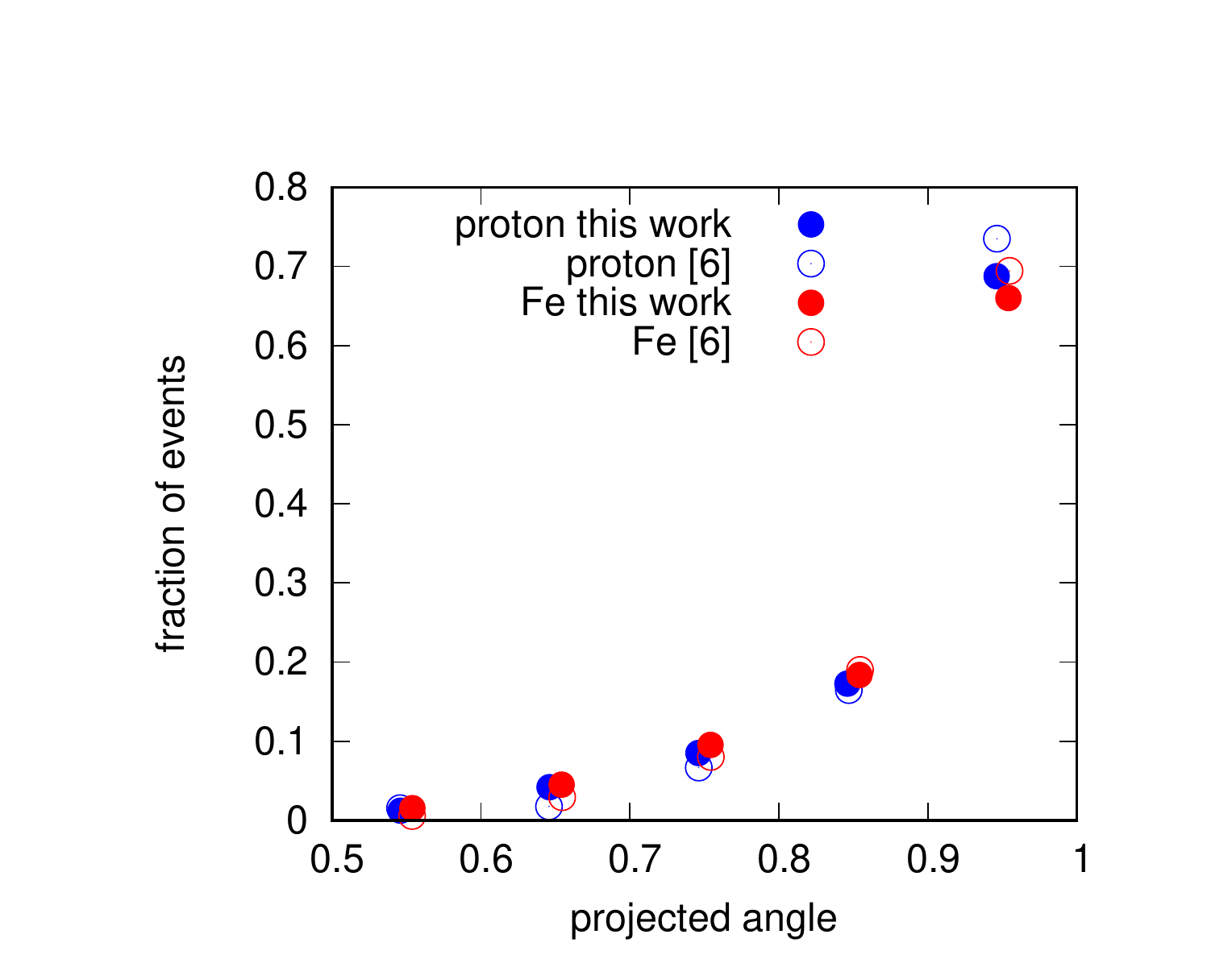}

\caption{Comparison of relative distribution of projected angle of muon bundles with observed multiplicity larger than 20 predicted by our simplified DELPHI model with simulated values from \cite{DELPHI} for proton and iron primaries. The values have been slightly shifted on the x-axis for readability.}
\label{theta}
\end{figure}

To avoid these issues, we use a different approach: from \cite{Travnicek} we know how the data compares with QGSJET-01 \cite{QGS01a,QGS01b} proton and iron simulation -- namely that for observed multiplicities larger than 20, there are $(2.24\pm0.17)$ times more events than in pure proton simulations and, for multiplicities larger than 80, there are $(1.45\pm0.23)$ times more events than in pure iron simulations; we denote these ratios $DPH_{20}$ and $DPH_{80}$ from now on and use them as benchmark observables to compare interaction models with that. To this end, we perform simulations with QGSJET-01 and use them as the basis for any further comparison. Ideally, we would like to find which model reproduces the $DPH_{20}$ and $DPH_{80}$ the best for a realistic mixture of primary particles (as discussed later) -- not just for an extreme and most likely unrealistic assumption of pure iron primary beam.

Our simplified DELPHI is a geometric model which mimics the shape and rough properties of the HCAL -- it has two parameters, one corresponding to the chance to miss a muon traversing the detection volume due to the combination of dead spaces and finite efficiency of the sensitive parts. The other parameter corresponds to HCAL granularity, which is related to the saturation that occurs when more muons pass through one streamer tube. These values are fixed by hand by tuning the dependence of reconstructed muon multiplicity on the true one against the relevant information from \cite{Travnicek}. Interestingly, such a simplified description of the detector reproduces the results of much more involved simulations very well, as illustrated in Fig~\ref{theta} where the distributions of projected angle (to the plane perpendicular to the beam line) of observed muon bundles in DELPHI are compared for our model and that used in \cite{DELPHI}. The rock overburden is taken into account as a simple cutoff at 52 GeV for vertical muons with simple geometric zenith-angle dependence -- as the amount of mass is quite substantial, the fluctuation in energy losses of muons over the track is very small. Each CORSIKA shower is used 100 times with a random core position within 150 meters of the detector, a distance chosen so that for any shower, at least a third of core positions produce no muons in the detector.

\begin{figure}

\includegraphics[width=\textwidth]{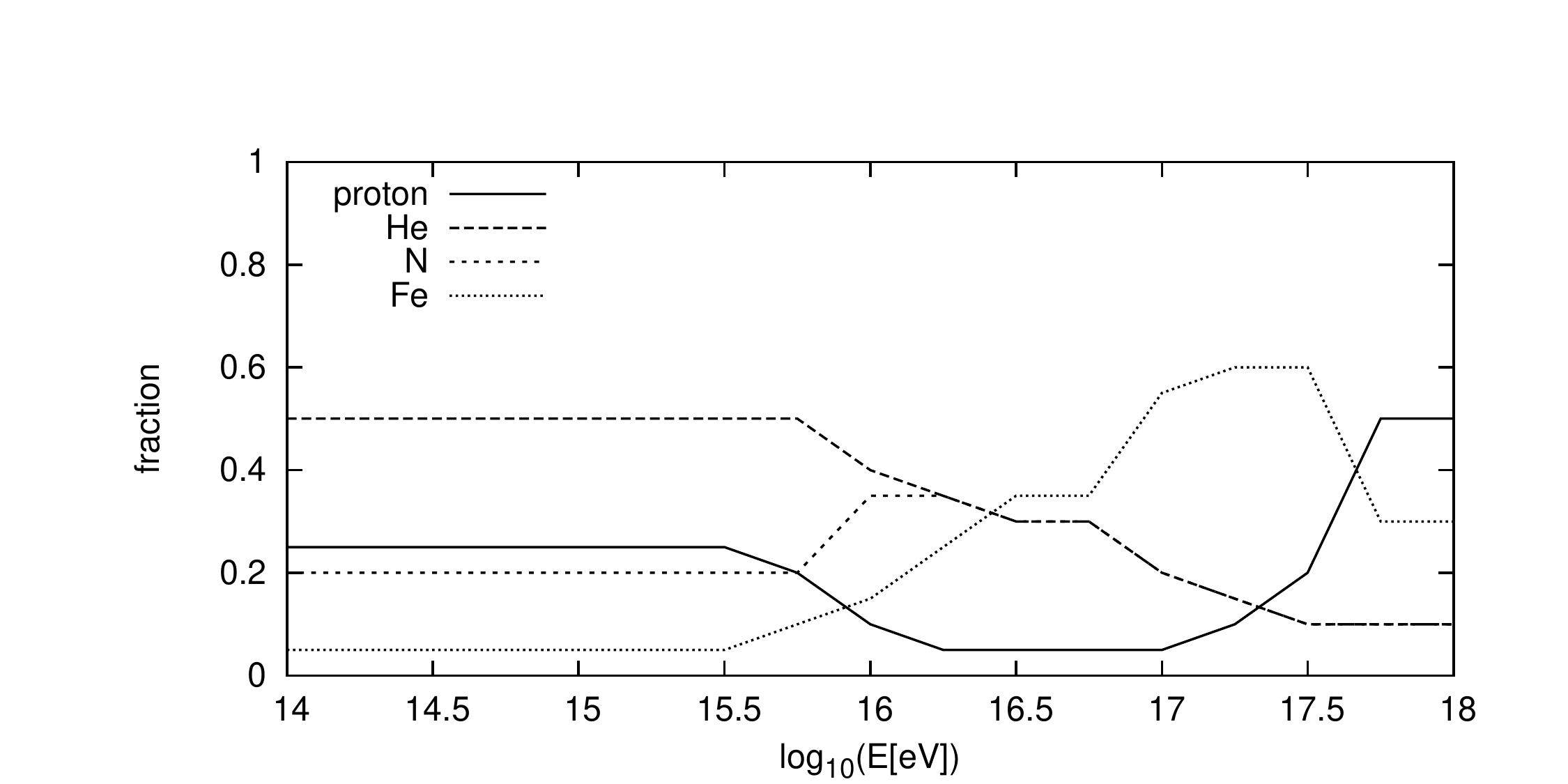}

\caption{Composition of the simulated primary beam in the relevant range of energies for DELPHI simulations. The first decade of energy has a negligible influence on the benchmark variables and is included mainly for consistency reasons.}
\label{fig:sloz}
\end{figure}

The CORSIKA\cite{Corsika} (version 7.37) simulations are carried out in the energy range of $10^{14}$--$10^{18}$ eV, which is sufficient for observed muon multiplicities above 10. The primary spectrum is simulated as $E^{-1}$ and re-weighted in the same way as in \cite{Travnicek} to $E^{-2.7}$ above the knee ($3\times10^{15}$ eV) and $E^{-3}$ below the knee. We do not have to make any assumptions about the overall primary flux, because we always compare simulations to simulations with the same overall number of events. However, it is interesting to note that the flux used in \cite{Travnicek} to derive the experimental values $DPH_{20}$ and $DPH_{80}$ was the maximal flux permitted by various cosmic ray experiments. Thus, any uncertainty in the flux only increases the observed effect. Varying the flux and spectrum in our simulations within the uncertainties quoted in \cite{KASCADE}  can change the $DPH_{20}$ and $DPH_{80}$  by up to 15~\% -- but only downwards. Four primaries (p, He, N and Fe) are simulated and a realistic energy-dependent primary beam is formed according to KASCADE and KASCADE-Grande data \cite{KASCADE} (Fig.~\ref{fig:sloz}).\footnote{Considering that in order to extract the primary composition from any EAS data, hadronic interaction models have to be used, we should be using a different composition for each model to ensure internal consistency of the procedure. While interpretations of KASCADE data for several different commonly used models are available, this is not the case for any modified models. However, there is consensus that below the knee the composition is rather light and then it gets heavier until reverting to mainly proton at $10^{18}$ eV according to the Auger data and this trend is reflected in our reference choice.} 

The simulated zenith angle is 0--60 degrees as inclined showers are highly suppressed due to the overburden increase with zenith angle. The low-energy model in CORSIKA is always set to be GHEISHA \cite{geisha}. We are aware of the shortcomings of this model, but we are consistently facing technical issues when using our modified interaction model (see sec.~\ref{spam}) together with FLUKA and we want to use the same setup for all simulations for the sake of comparison. Moreover, because of the overburden cut-off, the influence of the low-energy model on DELPHI simulations is very small. We found that the difference in the values of the benchmark variables (explained  below) between GHEISHA and FLUKA is roughly 2~\% when tested with unmodified QGSJET-II-04, where the use of FLUKA is possible.

Because we are only interested in muons, we do not follow EM particles in the CORSIKA simulations and the EGS option is turned off. We are thus losing some very small part of muons this way, but as we do the same for the reference simulations, this is a second order effect. Combined with the large energy threshold of underground detection and relatively low energy of most of the showers, we are able to do all simulations without thinning, conveniently avoiding the potential issue of interplay between the granularity of the detector and un-thinning. Because the muon multiplicity spectrum falls very quickly, we need relatively large statistics of showers to reliably cover the high-multiplicity part. Each run for a given model consists of 23775 showers and takes several months of CPU time which we are usually able to achieve within days on our local cluster. For unmodified QGSJET-II-04 we have performed 10 such runs, the RMS variance of the resulting set is 1 \% for $DPH_{20}$ and 0.7 \% for $DPH_{80}$. We have also performed simulations using each shower 1000 times instead of 100 times, again finding only a 1 \% difference.

In Tab.~\ref{tab:MCs} we show results of the DELPHI muon bundle simulation for different standard hadronic models (QGSJET-II-03, QGSJET-II-04 and EPOS LHC) as ratios with respect to QGSJET-01. For each model we show the $DPH_{20}$ ratio for pure proton, the $DPH_{80}$ ratio for pure iron and both ratios for our chosen mixed composition. Note that the latter values are obviously non-unity even for QGSJET-01 itself and that the discrepancy between data and simulations is even higher when it is taken into account that pure iron composition across the whole energy range is not plausible. We remind the reader that the ``target'' values are $DPH_{20}= 2.24\pm0.17$ and $DPH_{80}=1.45\pm0.23$.

\renewcommand{\arraystretch}{1.2}
\begin{table}
\centering
\begin{tabular} {c c c c c }
model & $DPH_{20}$ & $DPH_{80}$ & $DPH_{20}$ & $DPH_{80}$ \\
composition & p only & Fe only & mixed & mixed \\
\hline
QGSJET01 & 1.00 & 1.00 & 1.43 & 0.70 \\
QGSJET-II-03 & 1.11 & 0.75 & 1.54 & 0.57 \\
QGSJET-II-04 & 1.11 & 1.37 & 1.72 & 0.83 \\
EPOS-LHC & 0.85 & 0.86 & 1.27 & 0.59 \\
DELPHI data & – & – & 2.24 & 1.45 \\
\end{tabular}
\caption{Comparison of various Monte Carlo generators with DELPHI observations of muon bundles as ratios with respect to QGSJET-01.}
\label{tab:MCs}
\end{table}

From this table one can make several interesting observations. The first one is that the evolution from QGSJET-01 to QGSJET-II-03 has actually made the discrepancy worse for the highest-multiplicity region. The other observation is that while both new models are tuned to the same LHC data they show a large difference between them. That is particularly puzzling because the particles that contribute the most to multi-muon events interact at c.m.s.~energies not too different from those achieved at the LHC. This is another hint that if we want to take the muon excess at DELPHI at face value, we need to start thinking about modifying the models right at the LHC energies – obviously while making sure that we do not contradict the LHC data.

\section{Soft Particles in Cosmic Ray Showers}

\begin{figure}

\includegraphics[width=\textwidth]{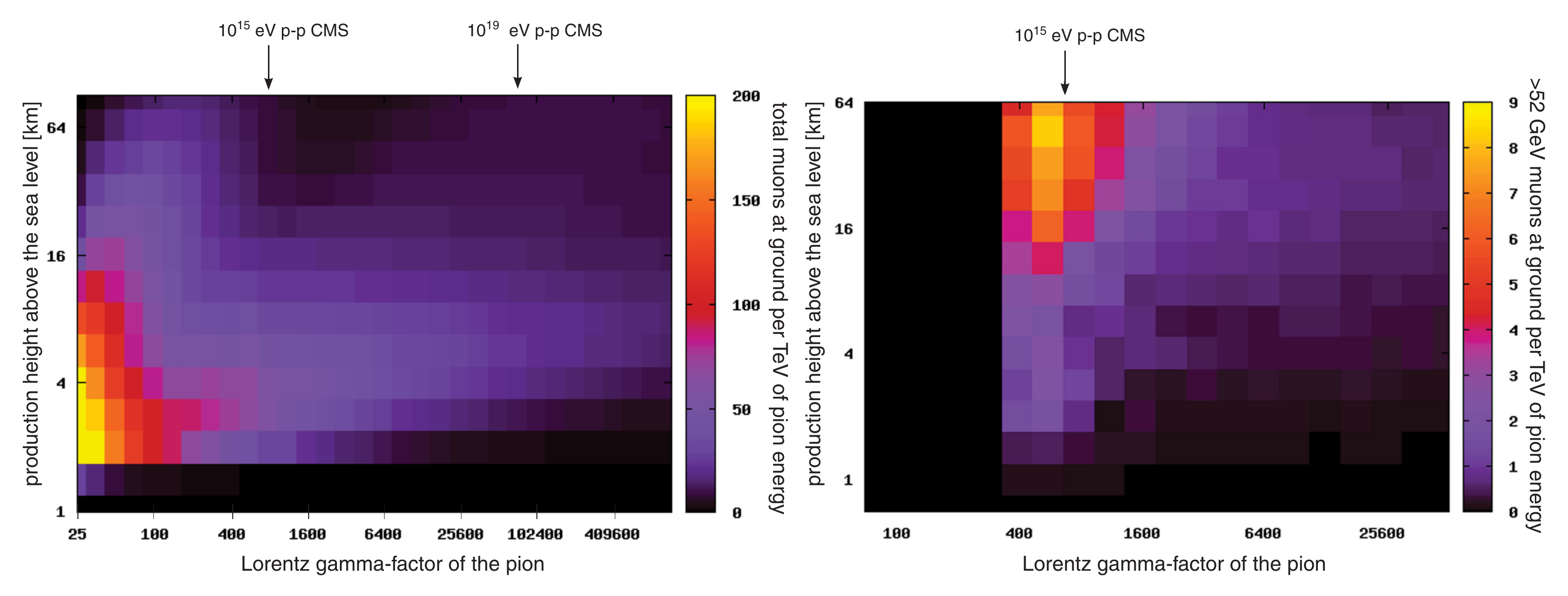}

\caption{The number of muons at ground produced per TeV of laboratory energy of a pion depending on its production height and gamma factor. Left panel shows all muons above 100 MeV, right panel shows only muons with energy greater than 52 GeV. Gamma factors of the center-of-mass systems for proton-nucleon interactions are indicated for orientation. }
\label{fig:vajicka}
\end{figure}

The muons observed at DELPHI have momentum cut-off of about 52 GeV for vertical muons imposed by detector overburden. Their momenta were studied in detail by L3+C and the unfolded surface flux peaks at around 100~GeV \cite{L3C}. We would like to asses which particles in the shower are most likely to create such high-energy muons and muons in general, motivated by the question if a possible source of such muons could be decays of soft pions produced in local c.m.s. along the shower development. To this end, consider a secondary charged pion created in some interaction in the shower with a given energy and height above the ground. This pion can either decay into a muon or interact and produce a hadronic subshower. In any case, by performing repeated simulations, starting always with a single pion, we can obtain the average number of muons at ground caused by such secondary pion and its potential daughter particles. However, this number will be largely proportional to the energy of the pion, as more energetic particles are more likely to create more secondaries, because of simple energy conservation. We can thus view the whole shower as having a budget, given by the energy of the primary particle. This approach is visualized in Fig.~\ref{fig:vajicka} where in the left panel we have counted the total number of muons of interest to the Pierre Auger Observatory. The threshold muon energy was therefore set to 100 MeV. In the right panel is the number of high-energy muons with energy above 52 GeV, capable of penetrating into the DELPHI cavern. In both cases, we plot this number divided by the energy of the initial pion in TeV. Note that in both cases, the ground level appropriate for the given experiment was chosen. Shown in this way, one can see the "muon efficiency" of particles with different initial conditions: if one were to modify the hadronic interaction models, the overall energy in the shower would have to stay conserved -- thus, if the goal is to increase the number of muons on the ground in the simulations, it is clearly necessary that particles that cause high number of muons at ground per unit of energy are added, not just particles that create a high number of muons, because there may be insufficient energy to support a large enough number of them. We quantify the energy of the pion by its Lorentz gamma factor and indicate the gamma factors of the center-of-mass frames of interactions of particles at specific energies. When we later add particles with very low momenta in these frames, they will naturally fall close to these indicated positions in this plot.

At the first sight, these results seem very different, but they are, in fact, highly complementary. Considering that the most important range of primary energies for the DELPHI data is around $10^{15}$--$10^{17}$ eV and that the first interactions of cosmic rays occur high up in the atmosphere, it is obvious that the largest efficiency in adding muons observable at DELPHI comes from pions produced in the first few interactions with small relative momenta with respect to the center-of-mass system of this interaction. On the other hand if one wants to produce more muons in simulations of showers detected by the Pierre Auger Observatory, it is desirable to produce pions deeper in the atmosphere and with small momenta. The very same energies at which first interactions occur in a $10^{15}$ eV shower high in the atmosphere correspond to a much later and deeper generation of particles for a $10^{19}$ eV shower. While it is difficult to access directly the yellow bottom-left corner of the left panel of Fig.~\ref{fig:vajicka} because the interactions there are governed by mostly known low-energy interaction physics and pion decay, one can still see that producing as many soft pions as possible close enough to ground can be beneficial. In fact, this is not the only interplay between DELPHI and Auger data and we will revisit this topic later.

\section{Soft Particle Addition Model and its Implementation}\label{spam}

The muon excess observed in \cite{Travnicek} must be produced at incident momenta of cosmic particles around ${p}_{\mathrm{Lab}} \approx 10^{15}$--$10^{17}$ eV -- a range, the lower part of which, corresponds to the LHC energies. However, no such effect has been detected by LHC experiments. These two facts could be reconciled provided that all corresponding soft particles are produced in remnant jets i.e. in a very narrow cone along the interaction axis. The detection threshold of LHCf is around $100$~GeV \cite{LHCf}, i.e. too high for this purpose. A place to observe increased production in remnant jets at sufficiently high energies are cosmic ray showers. In this case the increased pion production in the forward direction would increase the number of muons observed at ground.

To test the influence of the soft particle production in remnant jets on cosmic ray showers we have set up a simple model which adds soft particles (i.e. particles with very low momenta in the center-of-mass system of the collision) to the remnant jets in the forward-backward direction of the collision. The rest of the interaction is modeled by one of the standard Monte Carlo generators used in high energy cosmic ray physics.  We have tried to make a set of plausible simplifications to arrive at a concrete implementation which depends only on a handful of free parameters. Even so the complete parameter space is large and what we present is more an indication of the dependence of the results on each individual parameter than a comprehensive scan of this space. We stress that our ideas concern only non-perturbative phenomena in remnant jets which are much less understood compared to the jets originating from parton-parton scattering. 

While the theoretical motivation was very different, a specific implementation of adding soft particles to interactions has already been explored in the context of the formation of quark-gluon plasma with negative results concerning the DELPHI CR measurements \cite{QGP,Nosek2}. The likely reason for the differences of the results when compared with the present work lies not only in the precise values of the parameters of the implementations, but also in the fact that the QGP-like addition of soft particles was always performed only in the first interaction of the CR shower, whereas in this paper, we modify potentially a high number of interactions down to a certain energy threshold. On the other hand, the superficially big difference in geometry (isotropic for QGP vs. narrow cones for our model) has been found to have a rather small effect, because most of the soft particles we add have small proper momenta with respect to the boost of the c.m.s.  

Instead of adapting the code of an existing hadronic interaction model we use the list of secondaries produced by the QGSJET-II-04 interaction model and modify it before the particles are passed to CORSIKA. Thus, the majority of the features of the interactions are inherited from the base model. Most of the free parameters are configurable using custom CORSIKA steering keywords allowing easy tests of their effect on the simulation. While building the model, we have in mind not only the desire to increase the muon production to levels observed by DELPHI and the Pierre Auger Observatory, but we also take care not to become at odds with any experiment we are aware of. It turns out that Auger data on longitudinal shower development \cite{Xmax} provide a very strong constraint, in particular around $3\times 10^{18}$ eV where the observed values for the mean depth of shower maxima are very close to model predictions for protons. Like most mechanisms that add muons to UHECR showers, soft-particle production tends to decrease the predicted value of the depth of shower maximum. Clearly a shift of more than roughly 30 g\,cm$^{-2}$ would be incompatible with Auger data in this energy range.

In our simulations, we first determine the center-of-mass system of the nucleons that are actually interacting -- that is, when one of the participants is a nucleus, we take into account only the wounded nucleons as decided by QGSJET and ignore the spectator nucleons. From this point, we work in this reference frame, only transforming all energies and momenta to the laboratory frame at the end. The choice of the reference frame might have an impact on the results, in particular on the fluctuations of the depth of the shower maximum $X_\mathrm{max}$ shown in Sec.~\ref{Results}, because it varies from interaction to interaction based on the number of wounded nucleons in both the projectile and the target. Those results should thus be interpreted with caution.

In the next step, we assume that a certain fraction of the total energy $\sqrt{s}$ would be converted into soft particles. While our code allows for this fraction $f$ to be set as energy-dependent, all presented simulations use, for simplicity, a sharp energy threshold above which $f$ is constant. The efficiency of the process can be the same for all collisions, or it can depend on the size of the interacting system. This dependence is expressed by modifying the fraction $f$ by the sum of the numbers of wounded nucleons in the projectile and in the target raised to a power $\eta$. We will present results for various choices of $\eta$ both positive, negative and also $\eta=0$ and show that the sensitivity to this choice is rather small in all observables considered apart of the longitudinal development of iron-induced showers at very high energies. 

In order to add any particles to the interaction, we must make room for them so that energy and momentum are conserved. One option is to remove some particles, but that could potentially affect many other features of the interaction and we would have to account for baryon number conservation, baryon/meson ratio for leading particles etc. Thus, we keep all the produced particles, but remove some fraction of their momenta, while keeping the directions of the momenta unchanged. It turns out that removing the same fraction of momentum from all particles is actually detrimental to the predictions for DELPHI. This is not surprising, because then energy is also taken from already very soft particles. It also has tendency to change the predicted $X_\mathrm{max}$ in the energy range of interest for the Pierre Auger Observatory a lot before even considering any additional particles. Instead, we select only particles with energy above 10 GeV in the c.m.s frame where we work and on top of that, the momentum fraction taken from each particle is proportional to the logarithm of its energy. We have generated millions of showers by QGSJET-II and in all cases we have found particles suitable to be modified. We also remove energy only from those particles whose momenta are within specific cones along the collision axis -- those are the same in which we are going to add the soft particles. We never change the directions of the momenta of the particles in the c.m.s. The overall normalization of the momentum removal is fixed by the requirement of the overall energy conservation in an iterative procedure.

We have found during the review that this procedure is not perfectly exact when viewed from the laboratory frame. For a test set of 1000 proton induced interactions at $10^{17}$ eV we have found the mean imbalance in longitudinal momentum in the laboratory frame to be $-0.6$ \% of the initial momentum with RMS of $1.7$ \% while transverse momentum is conserved perfectly within the precision of the calculation. This means that on average a small fraction of the primary energy goes missing. The results for iron induced showers are much the same (mean $-0.8$ \%, RMS $2.1$ \%) . As the amount of muons in the shower increases with primary energy, this effect could possibly lower the amount of muons predicted by our simulations, which has to be kept in mind when interpreting our results.

The next step is the actual addition of particles. To this end we must choose the probability to generate particles of different species, the opening angle from the collision axis and the momentum distribution of the particles. Here we use only the most ubiquitous particles -- pions, kaons and nucleons -- and the ratios between these types are taken from typical ALICE observations at the LHC \cite{ALICE} as a contemporary example of hadronization in a high-energy environment. The isospin symmetry is strictly upheld, meaning 2/3 of pions are charged and 1/3 are neutral pions and similarly for $K$ and nucleons. The particles are always added as a back-to-back pair of particle-antiparticle so that all other conservation laws are also naturally satisfied. 

The opening angle from the collision axis is in principle a free parameter, but both the theoretical motivation and possible detection constraints at accelerators dictate that it is small -- as noted already above, soft particles are essentially invisible in the forward direction at the LHC\footnote{If we were to postulate soft-particle addition in the scope needed for observable effects in CR data to take place at small rapidities, the LHC detectors would have tested (and disproved) the model immediately, because the central detectors are generally sensitive well below 1 GeV and the number of particles added in our model can be very large, as illustrated in Sec.~\ref{Results}.} We set the angle to 1 degree in all the simulations -- only during the review of the article we were made aware of the work \cite{TOTEM} done with the TOTEM detector at the LHC, which would already disprove such a model for most of our parameter space. In \cite{TOTEM}, it is stated that the reconstruction efficiency is 90~\% for $p_\mathrm{T}>20$~MeV -- even if it dropped to zero immediately, we expect roughly 4--10 extra particles detected in TOTEM, clearly at odds with their measurement. This number however quickly decreases when narrowing the opening angle in our model -- already at 0.6 degrees we get only $\approx0.5$ particles on average and at 0.1 degrees the added particles are invisible to TOTEM. While we have made all the simulations at 1 degree, we have tested several parameter choices with the opening angle of 0.1 degrees and the results are consistent within statistical errors.

 As for the shape of the momentum distribution, we have considered $\exp(-p/p_{0})$, $p\exp(-p/p_{0})$ or
$p^{2}\exp(-p/p_{0})$ -- if we assume that the momentum distribution is exponential, the first and the last choices represent the two extremes of how the distribution limited to a narrow cone
is generated -- whether there is a cut-off in the production angle or it is compressed, while the middle distribution (used henceforth) represents a compromise option. The differences between the distributions are of the order of 3--7~\% in the predicted frequency of high-multiplicity muon bundles, much smaller than the effect of other parameters.

Finally, the parameter of the momentum distribution $p_{0}$ has to be chosen (for the $p\exp(-p/p_{0})$ distribution we use, the mean particle momentum is $2p_{0}$). As we will illustrate later, the muon gain depends mainly on the numerical quantity of particles added and thus it decreases when more energy is used to gain kinetic energy of the products instead of being converted to their mass. However, one can easily argue that in a strongly interacting medium the typical momenta of particles will be at least of the order of 200 MeV. The observation can be stated more precisely based on the results of Ref.~\cite{Ridky}, where it is shown that the characteristic $p_T$ distribution of pions over a wide range of energies behaves as $\exp(-m_T/a)$ where $a\approx170$ MeV and $m_T$ is the transverse mass. Incidentally, for pions, the best approximation for this distribution using an exponential function in $p$ is such that $p_0\approx200$ MeV . Aware that the kinematic conditions at hand are quite different, but inspired by the wide applicability of this value, we use $p_0=200$ MeV as the basic choice in our simulations. For comparison, we present simulations where $p_{0}=200$ MeV at the energy threshold for the soft-particle addition and grows logarithmically with c.m.s. energy to 500 MeV at 100 TeV (we denote this choice of $p_0$ as "200/500" for brevity). The dependence of the results on the choice of $p_{0}$ is strong, as will be briefly shown later after properly introducing the DELPHI observables. Note that all these choices of $p_{0}$ lead, together with the small opening angle, to the production of particles for which the $p_{\mathrm{T}}$ distribution is sharply peaked at zero and significantly different from that measured in centrally produced particles.

\section{Simulations}\label{Results}
\subsection{Setup and Variables}

In addition to simulations for DELPHI, which we have already described, we have also conducted simulations with proton and iron primaries at the energy of $3.2\times 10^{18}$ eV with the results of the Pierre Auger Observatory in mind. For each relevant combination of input parameters we simulate 1000 showers for each primary particle and from these sets we extract the mean depth of the shower maximum $\langle X_\mathrm{max}\rangle$ by fitting the simulated longitudinal profile of each shower with a 4-parameter Gaisser-Hillas \cite{Gaisser-Hillas} function and express the result as an offset with respect to pure QGSJETII-4 simulations. Similarly, the number of muons at ground is extracted from the CORSIKA output without any detector simulation: for each shower, the NKG \cite{NK},\cite{G} function is fitted to the numerical muon density between 250 and 1500 meters from the core and the value at 1000 meters (corresponding to a typical Auger observable) is evaluated; the resulting mean number of muons $\langle N_{\mu}\rangle$ over each set of showers is in this case reported as a ratio with respect to pure QGSJETII-4 simulations. The thinning was set to $10^{-5}$ for the hadronic part of the shower and $10^{-3}$ for electromagnetic particles and the magnetic field and altitude corresponding to the location of the Pierre Auger Observatory were used.

Together, we consider five observables. From simulations at energies between $10^{14}$--$10^{18}$ eV we extract the two DELPHI benchmark variables $DPH_{20}$ and $DPH_{80}$ as explained in Sec.~\ref{delphi}. From simulations for both proton and iron primaries at $3.2\times 10^{18}$ eV we extract the mean atmospheric depth of the maximum of shower development $\langle X_\mathrm{max}\rangle$, its mean variance $\langle\mathrm{\sigma}(X_\mathrm{max})\rangle$ and the mean number of muons at ground at 1000 meters from the shower core $\langle N_{\mu}\rangle$. These five observables are presented as a function of four main parameters: the energy fraction $f$ converted in each interaction to soft particles, the dependence on the size of the interaction system expressed as the exponent $\eta$, the choice of the $p_{0}$ parameter in the momentum distribution and the energy threshold above which the soft-particle addition takes effect. In the rest of this section we present some views of the results of simulations equivalent to several centuries of CPU time.

An important technicality affects all the plots where the x-axis represents the converted energy fraction $f$: for non zero $\eta$ the actual fraction of energy converted to soft particles varies from interaction to interaction and the model parameter $f$ itself is not a good measure of the amount of energy used for particle production. Instead of $f$ we thus use the effective converted energy fraction $f_\mathrm{eff}$,  the weighted average of the actual fraction over all interactions within each shower, further averaged over all relevant simulated showers. For all simulations we weight the value by $\sqrt{s}$ of the interaction; additionally, for DELPHI simulations, we apply a correction for the varying contribution of different simulated showers to the resulting multiplicity spectrum as we are using one simulated shower 100 times with different core positions and we take into account the mass composition.

\subsection{Interpretation of the Plots}\label{plots}

Across the plots on Figs.~\ref{dcq}--\ref{rmsFe}, we use a consistent system of identification of the different simulations -- in general the shape, color and fill of the symbols represent the energy threshold (in $\sqrt{s}$), the choice of $\eta$ and of $p_0$ respectively as indicated by the brief legends next to each plot. The interpretation of the choice of threshold and $\eta$ is straightforward. As for $p_0$, the majority of simulations use two main choices: constant $p_0=200$ MeV (shown in empty symbols) and "200/500" ($p_0=200$ MeV at threshold, logarithmically increasing to 500 MeV at c.m.s energy of 100 TeV, filled symbols). Additionally, Figs.~\ref{dcq}, \ref{np}, \ref{nFe} and \ref{mux} include simulations with $p_0=800$ MeV and 3 TeV. For those $\eta=1/3$ and the energy threshold is 1 TeV. For the convenience of the reader we note that the inclusion or exclusion of these additional simulations is the only difference among the legends of Figs.~\ref{np}--\ref{rmsFe}; only Fig.~\ref{dcq} has a significantly different legend as it uses only a small subset of the simulations.

When the DELPHI-related benchmarks $DPH_{20}$ and $DPH_{80}$ are shown (Figs.~\ref{p20}, \ref{Fe80}, \ref{xp} and \ref{xFe}), a horizontal line indicates the observed value from \cite{Travnicek} with a gray band representing one standard deviation quoted therein. The scale of the y-axis is always chosen so that it starts at the value given by unmodified QGSJET-II-4 simulations. Such simulations would thus lie exactly in the bottom left or bottom right corner of the plots, depending on the direction of the x-axis.

The high-energy observables $\langle X_\mathrm{max}\rangle$ and $\langle N_{\mu}\rangle$ are always presented relative to unmodified QGJSETII-4 simulations as $\langle\Delta X_\mathrm{max}\rangle$ and $\langle \Delta N_{\mu}\rangle$ respectively, while for $\langle\mathrm{\sigma}(X_\mathrm{max})\rangle$, the reference QGSJETII-4 value is just shown.

\subsection{Results}

\begin{figure}

\includegraphics[width=\textwidth]{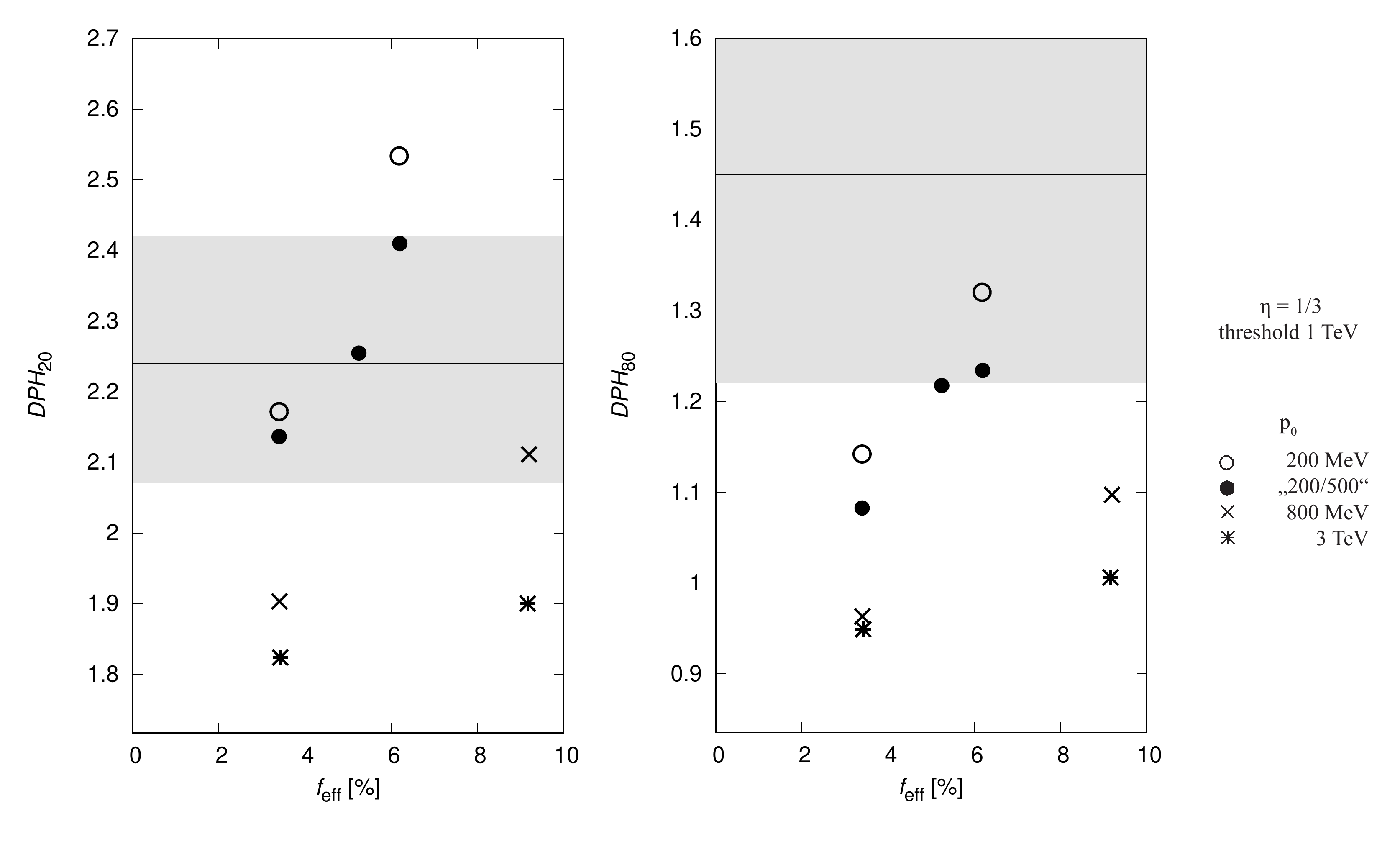}

\caption{$DPH_{20}$ (left) $DPH_{80}$ (right) as a function of the effective converted energy fraction $f_\mathrm{eff}$ for simulations with different choices of the mean momentum $p_{0}$ of added particles. Throughout all these simulations, the energy threshold was set to 1 TeV and $\eta=1/3$. The horizontal line shows the value measured by \cite{Travnicek} with a 1-sigma band. For more details on the plots see Sec.~\ref{plots}}
\label{dcq}
\end{figure}

Firstly, we show in Fig.~\ref{dcq} the strong dependence of the results on the mean momentum of added particles (which is $2p_{0}$ as mentioned in Sec.~\ref{spam}). As expected, the effect is strongest for the smallest momenta that correspond to the highest multiplicities. This plot supports our choice of the lowest reasonable value for this parameter.

\begin{figure}

\includegraphics[width=\textwidth]{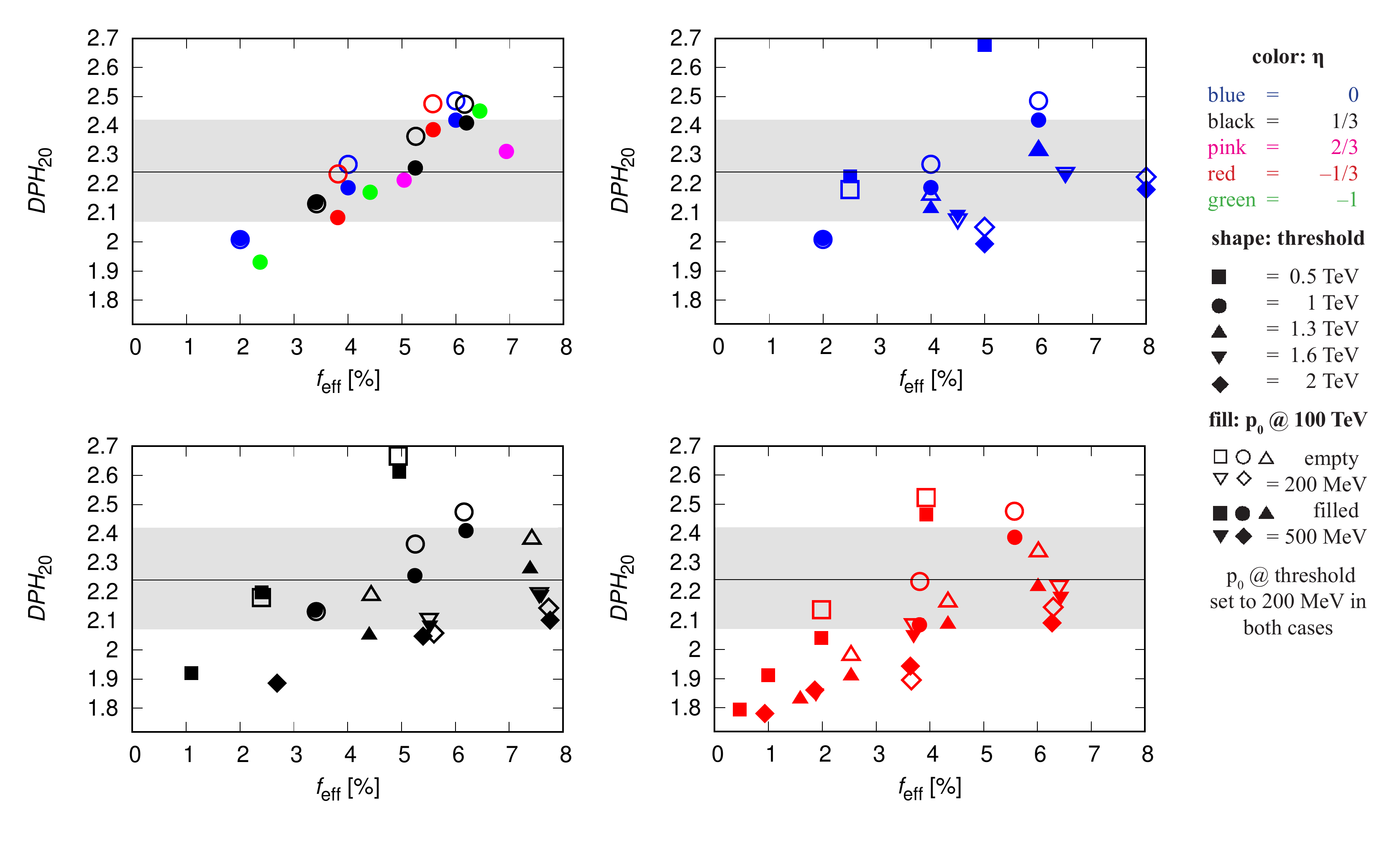}

\caption{$DPH_{20}$ as a function of the effective converted energy fraction $f_\mathrm{eff}$. The upper left panel shows the dependence on $\eta$ for a fixed energy threshold of 1 TeV, the rest show the dependence on the energy thresholds for particular choices of $\eta$. Both choices of $p_ 0$ are plotted. The horizontal line shows the value measured by \cite{Travnicek} with a 1-sigma band. For more details on the plots see Sec.~\ref{plots}}
\label{p20}
\end{figure}

\begin{figure}

\includegraphics[width=\textwidth]{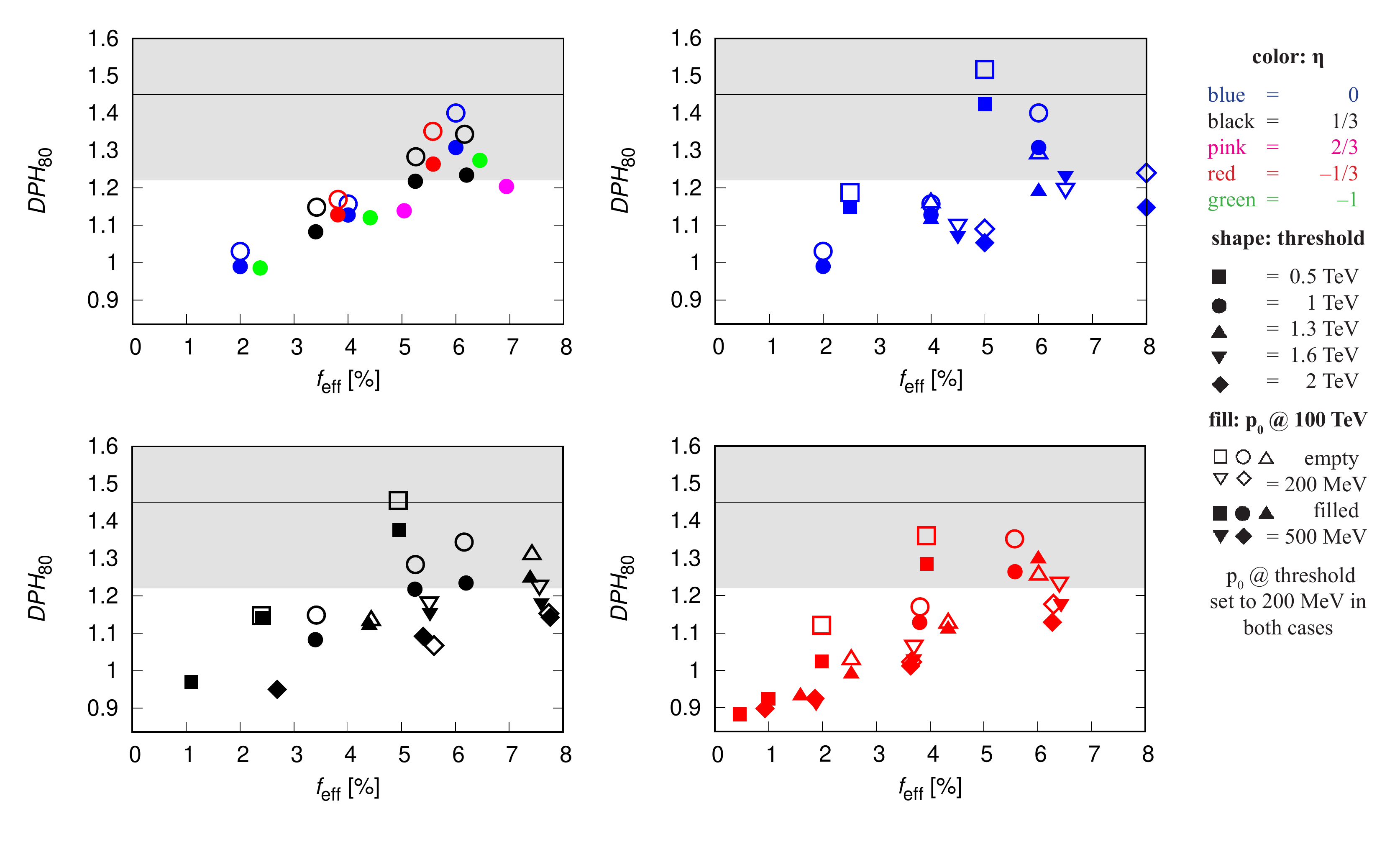}

\caption{$DPH_{80}$ as a function of the effective converted energy fraction $f_\mathrm{eff}$. The upper left panel shows the dependence on $\eta$ for a fixed energy threshold of 1 TeV, the rest show the dependence on the energy thresholds for particular choices of $\eta$. Both choices of $p_ 0$ are plotted. The horizontal line shows the value measured by \cite{Travnicek} with a 1-sigma band. For more details on the plots see Sec.~\ref{plots}}
\label{Fe80}
\end{figure}

Figs.~\ref{p20} and \ref{Fe80} show the results for the DELPHI benchmarks from simulations in the $f_\mathrm{eff}$-threshold-$\eta$-$p_{0}$ space -- that is first for simulations at fixed threshold of 1 TeV and various $\eta$ and then for three particular choices of $\eta$, each with energy thresholds varied between 0.5 and 2 TeV. It may seem that none of the models can completely describe the DELPHI data across all multiplicities within the assumed spectrum and composition, because while the target value for $DPH_{20}$ is reached at intermediate energy fractions, the target value of $DPH_{80}$ is reached only for rather high fractions or very low thresholds. However, when the experimental uncertainties of the DELPHI data (gray bands) are taken into account, the consistency is much better -- there are many points lying in the uncertainty bands simultaneously for both $DPH_{20}$ and $DPH_{80}$. 

In general, we observe that adding any soft particles, even at low fractions, increases the number of predicted high-multiplicity events. The dependence on the choice of $\eta$ is weak\footnote{Note that as the real converted energy fraction is calculated from the results of the simulations for non-trivial choices of $\eta$, the input parameters are estimated by averaging, hence the spread of the results over the horizontal axis which makes direct comparison slightly harder.}, while the dependence on the energy threshold is pronounced. For higher thresholds than 2 TeV, the target values are basically never reached and such simulations are thus omitted in favor of clarity of the plots; the required energy fraction gets lower for small thresholds, but lowering the threshold even further would go against the existing accelerator data notably those from fixed target experiments.

\begin{figure}

\includegraphics[width=\textwidth]{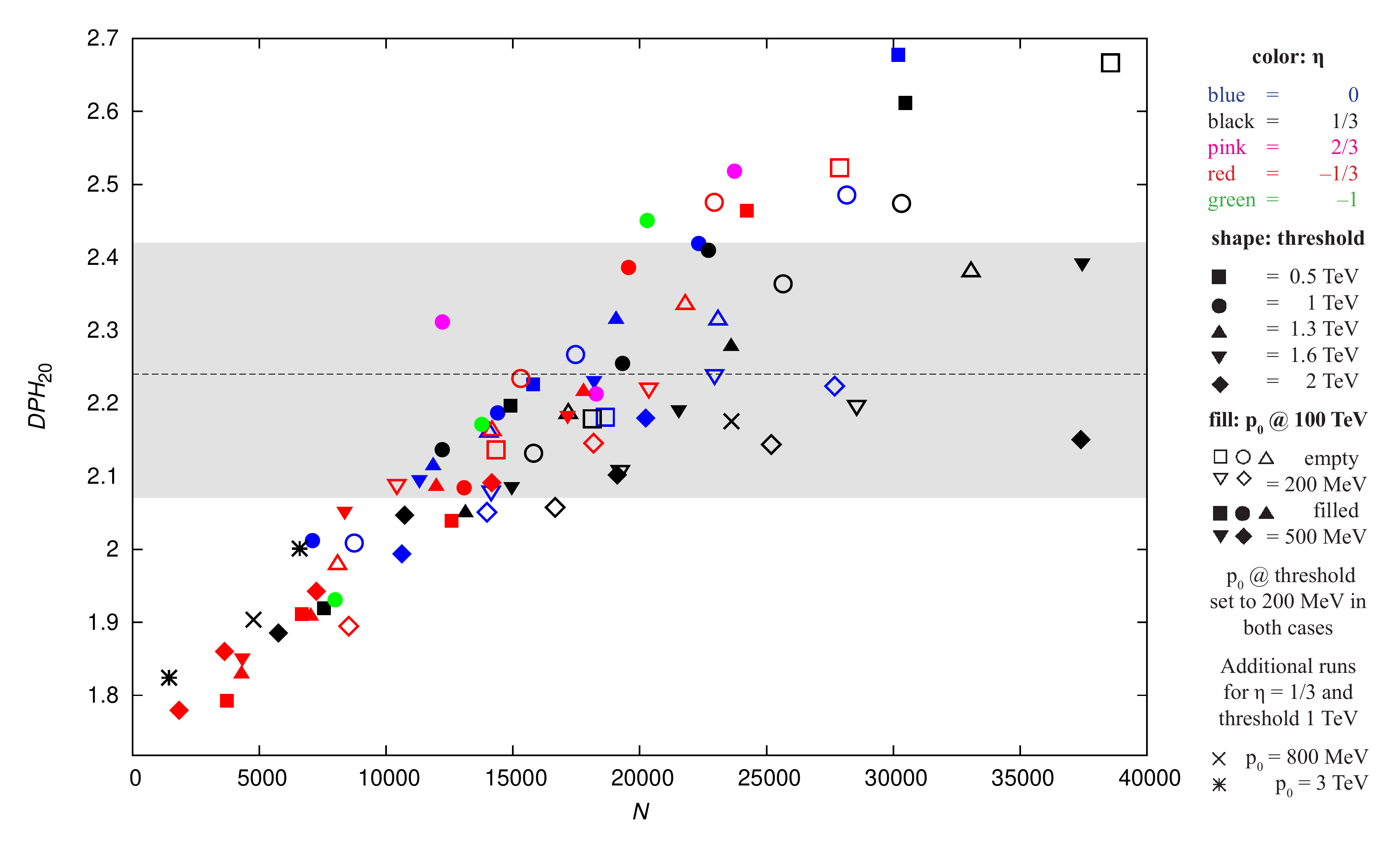}

\caption{$DPH_{20}$ as a function of the total number of soft particles $N$ added in one shower, averaged over all the simulated showers according to the chosen spectrum and mass composition. The plot includes all available simulations at once. The horizontal line shows the value measured by \cite{Travnicek} with a 1-sigma band. For more details on the plots see Sec.~\ref{plots}}
\label{np}
\end{figure}

\begin{figure}

\includegraphics[width=\textwidth]{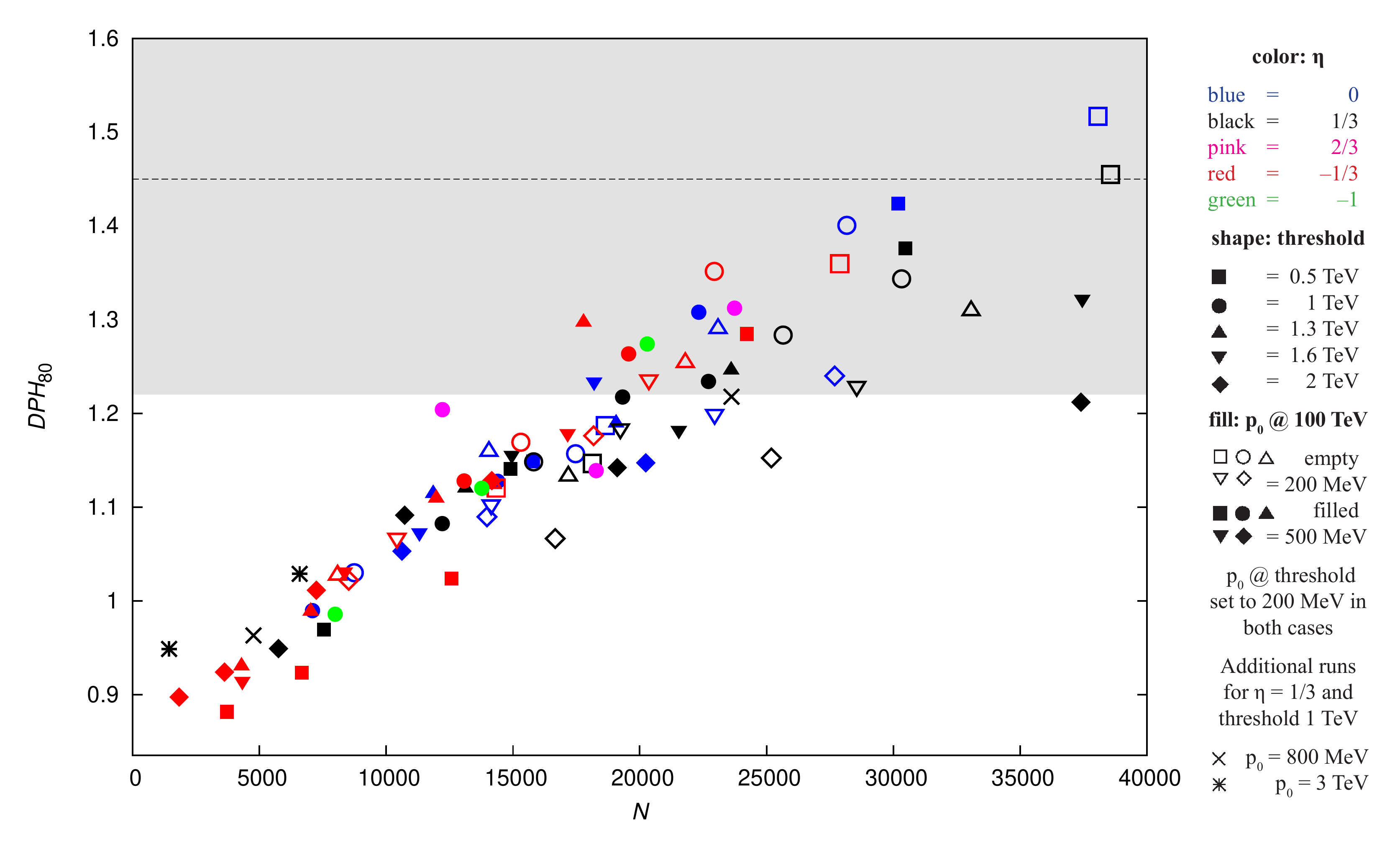}

\caption{$DPH_{80}$ as a function of the total number of soft particles $N$ added in one shower, averaged over all the simulated showers according to the chosen spectrum and mass composition. The plot includes all available simulations at once. The horizontal line shows the value measured by \cite{Travnicek} with a 1-sigma band. For more details on the plots see Sec.~\ref{plots}}
\label{nFe}
\end{figure}

Figs.~\ref{np} and \ref{nFe} show the same variables as a function of the total number of soft particles $N$ added in one shower, averaged over all the simulated showers according to the chosen spectrum and mass composition. This relation exhibits a very strong correlation for almost all choices of energy threshold, $\eta$ and even the mean momentum of added particles $p_{0}$, demonstrating that these parameters essentially only provide different proxies to the important underlying value of $N$. This we consider to be an important observation showing that while there are many choices that must be made when implementing a particular model, the general features of adding soft particles to the interactions do not strongly depend on these details. Similar observations have been made for general hadronic interaction models \cite{Ulrich}.

\begin{figure}

\includegraphics[width=\textwidth]{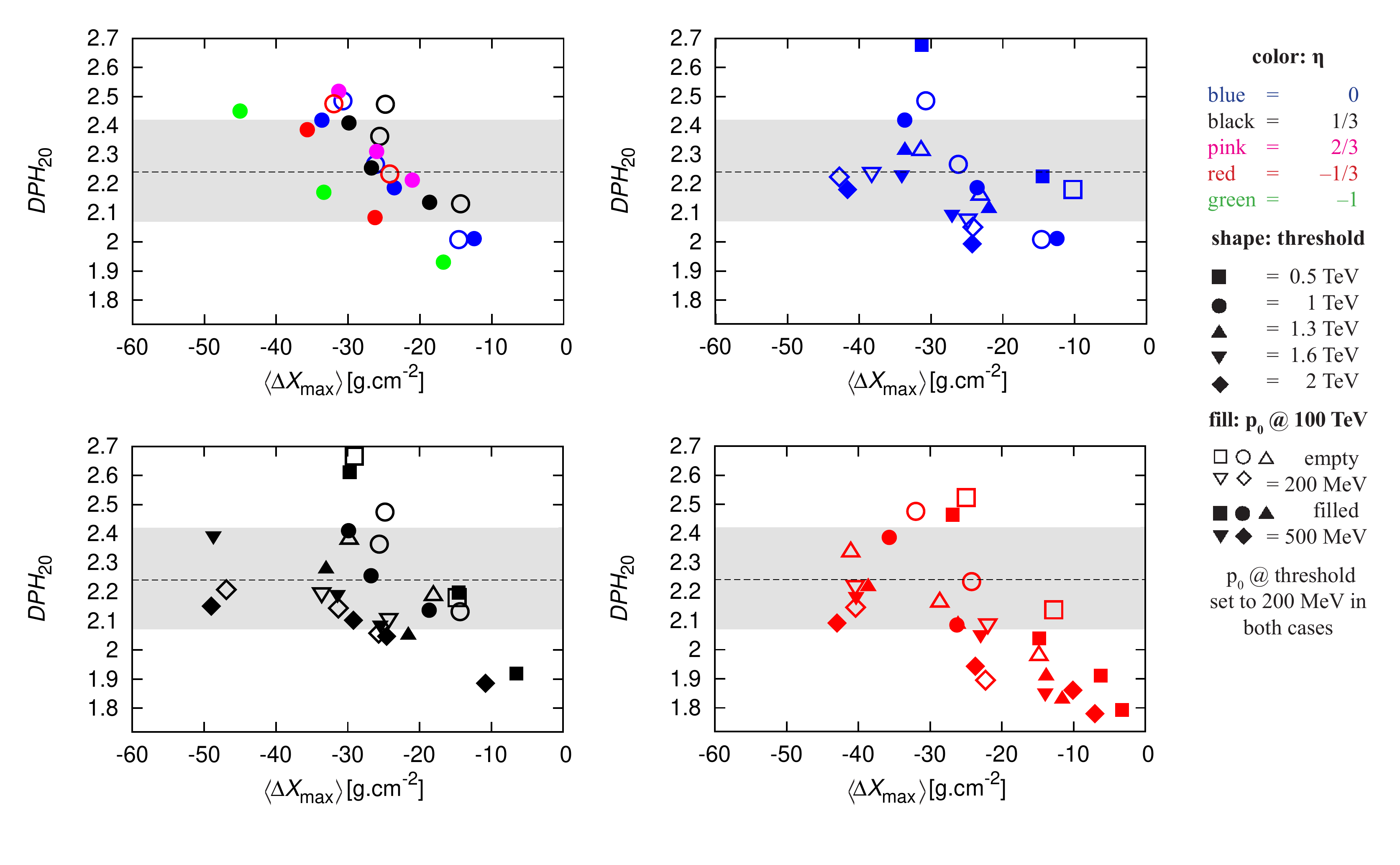}

\caption{$DPH_{20}$ as a function of $\langle\Delta X_\mathrm{max}\rangle$ of proton showers simulated with the same model parameters at $3.2\times 10^{18}$ eV. The upper left panel shows the dependence on $\eta$ for a fixed energy threshold of 1 TeV, the rest show the dependence on the energy thresholds for particular choices of $\eta$. Both choices of $p_ 0$ are plotted. The horizontal line shows the value measured by \cite{Travnicek} with a 1-sigma band. For more details on the plots see Sec.~\ref{plots}}
\label{xp}
\end{figure}

\begin{figure}

\includegraphics[width=\textwidth]{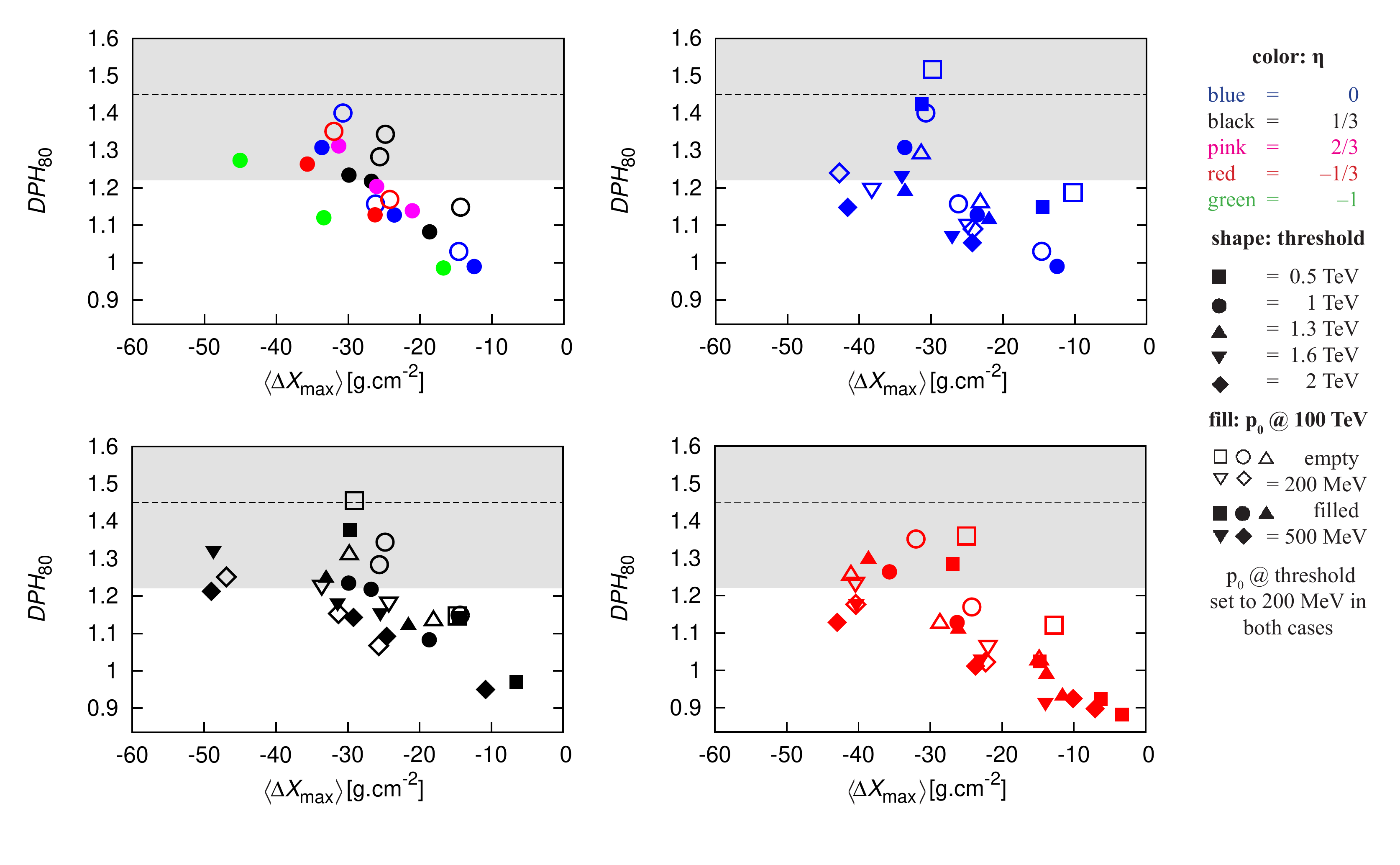}

\caption{$DPH_{80}$ as a function of $\langle\Delta X_\mathrm{max}\rangle$ of proton showers simulated with the same model parameters at $3.2\times 10^{18}$ eV. The upper left panel shows the dependence on $\eta$ for a fixed energy threshold of 1 TeV, the rest show the dependence on the energy thresholds for particular choices of $\eta$. Both choices of $p_ 0$ are plotted. The horizontal line shows the value measured by \cite{Travnicek} with a 1-sigma band. For more details on the plots see Sec.~\ref{plots}}
\label{xFe}
\end{figure}

Now we turn to the crucial interplay between the data from DELPHI and the Pierre Auger Observatory. As described in Sec.~\ref{spam}, the Auger data limit the acceptable amount of a shift downwards in the model predictions for the depth of the shower maximum around $3 \times 10^{18}$ eV to at most roughly 30 g\,cm$^{-2}$. From Figs.~\ref{xp} and \ref{xFe} we can see that some choices of parameters are more efficient in using the available wiggle room in longitudinal depth to generate muon bundles at DELPHI than others -- in particular the choices of large momentum parameter $p_{0}$ and large energy threshold are disfavored by Auger data as explanations to the DELPHI excess. As much as they already seemed less favored because the large fraction of converted energy needed to reach the target values for $DPH_{20}$ and $DPH_{80}$, it is the Auger data that gives concrete experimental evidence against the validity of those particular models at the ultra-high energies.

\begin{figure}

\includegraphics[width=\textwidth]{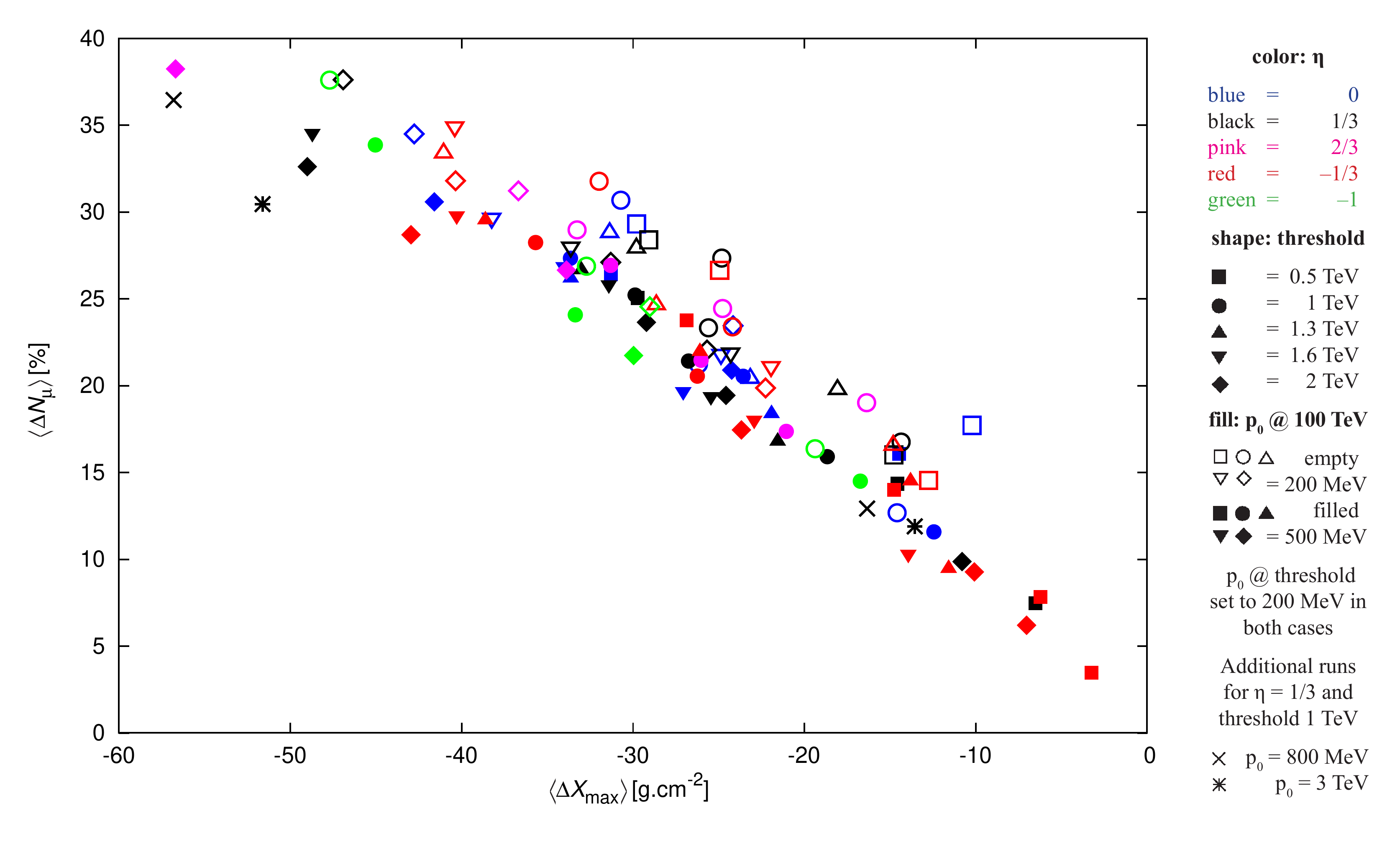}

\caption{$\langle\Delta N_{\mu}\rangle$ as a function of $\langle\Delta X_\mathrm{max}\rangle$ for proton showers at $3.2\times 10^{18}$ eV. The plot includes all available simulations at once. For more details on the plots see Sec.~\ref{plots}}
\label{mux}
\end{figure}

As we have already noted, there seems to be a muon excess in the UHECR data themselves. To this end, Fig.~\ref{mux} shows that the increase in the number of muons in proton shower with primary energy $3.2\times 10^{18}$ eV at 1000 meters from the shower core (a relevant measure of muon content for the Pierre Auger Observatory, extracted from the simulations as explained above) after the addition of soft particles strongly correlates with the shift in $\langle X_\mathrm{max}\rangle$ for various model parameters. Depending on the particular set up of the model, the muon gain when the shift is $-30$ g\,cm$^{-2}$ varies between 20 and 30 \%, which is in the ballpark of the unexplained excess reported several times by the Auger Collaboration \cite{AUGERmu1,AUGERmu2,AUGERmu3}. Note that the correlation between $\langle N_{\mu}\rangle$ and $\langle X_\mathrm{max}\rangle$ does not depend much on $\eta$ and energy threshold as the points tend to line up along a line, but the difference between simulations for $p_0=200$ MeV and for "200/500" is prominently visible, with the former providing consistently more muons for a given shift in $\langle X_\mathrm{max}\rangle$.

\begin{figure}

\includegraphics[width=\textwidth]{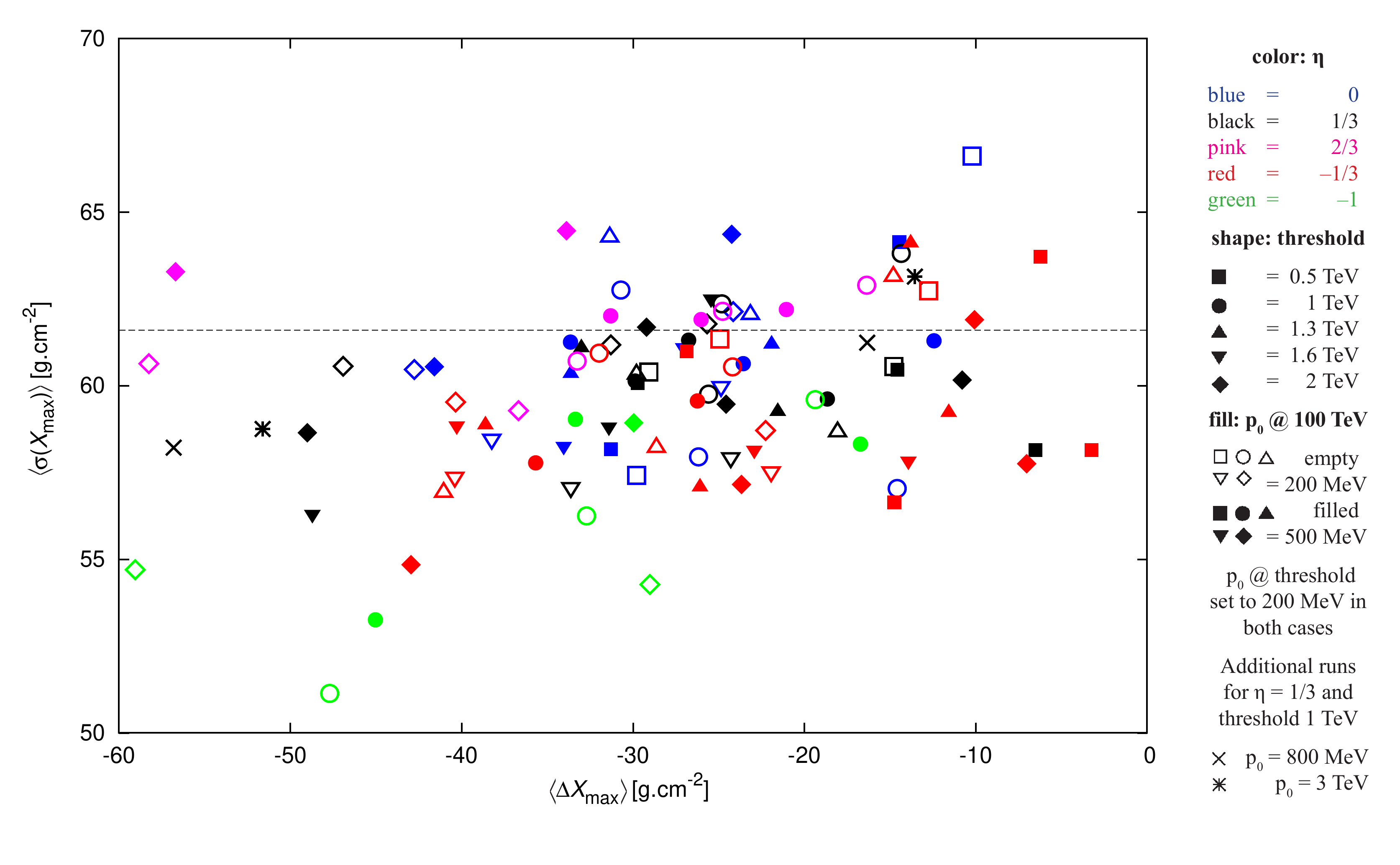}

\caption{$\langle\mathrm{\sigma}(X_\mathrm{max})\rangle$ as a function of $\langle\Delta X_\mathrm{max}\rangle$ for proton showers at $3.2\times 10^{18}$ eV. The plot includes all available simulations at once. The horizontal line corresponds to $\langle\mathrm{\sigma}(X_\mathrm{max})\rangle$ in unmodified proton QGSJETII-4 simulations. For more details on the plots see Sec.~\ref{plots} }
\label{rms}
\end{figure}

\begin{figure}

\includegraphics[width=\textwidth]{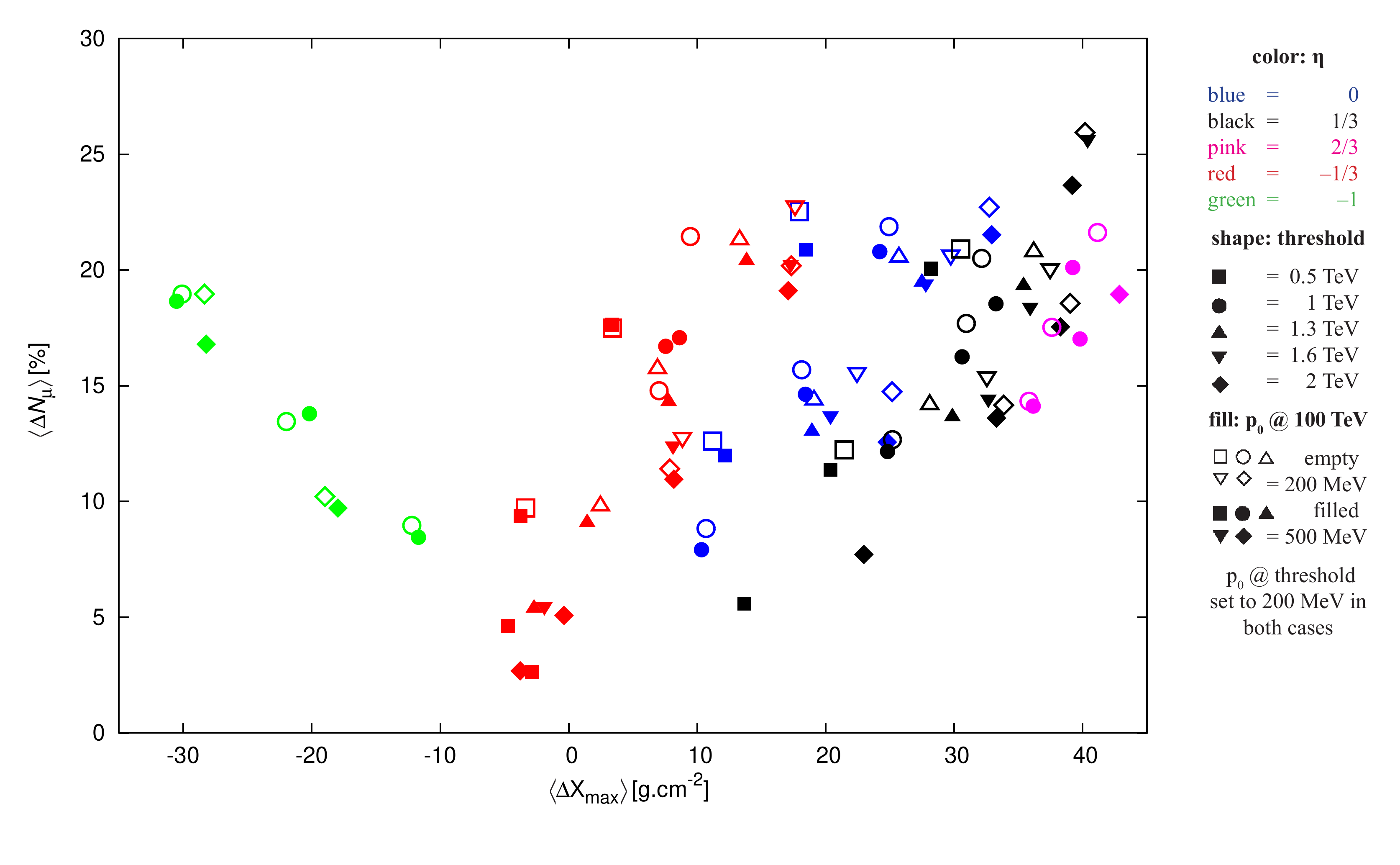}

\caption{$\langle\Delta N_{\mu}\rangle$ as a function of $\langle\Delta X_\mathrm{max}\rangle$ for iron showers at $3.2\times 10^{18}$ eV. The plot includes all available simulations at once. For more details on the plots see Sec.~\ref{plots}}
\label{muxFe}
\end{figure}

\begin{figure}

\includegraphics[width=\textwidth]{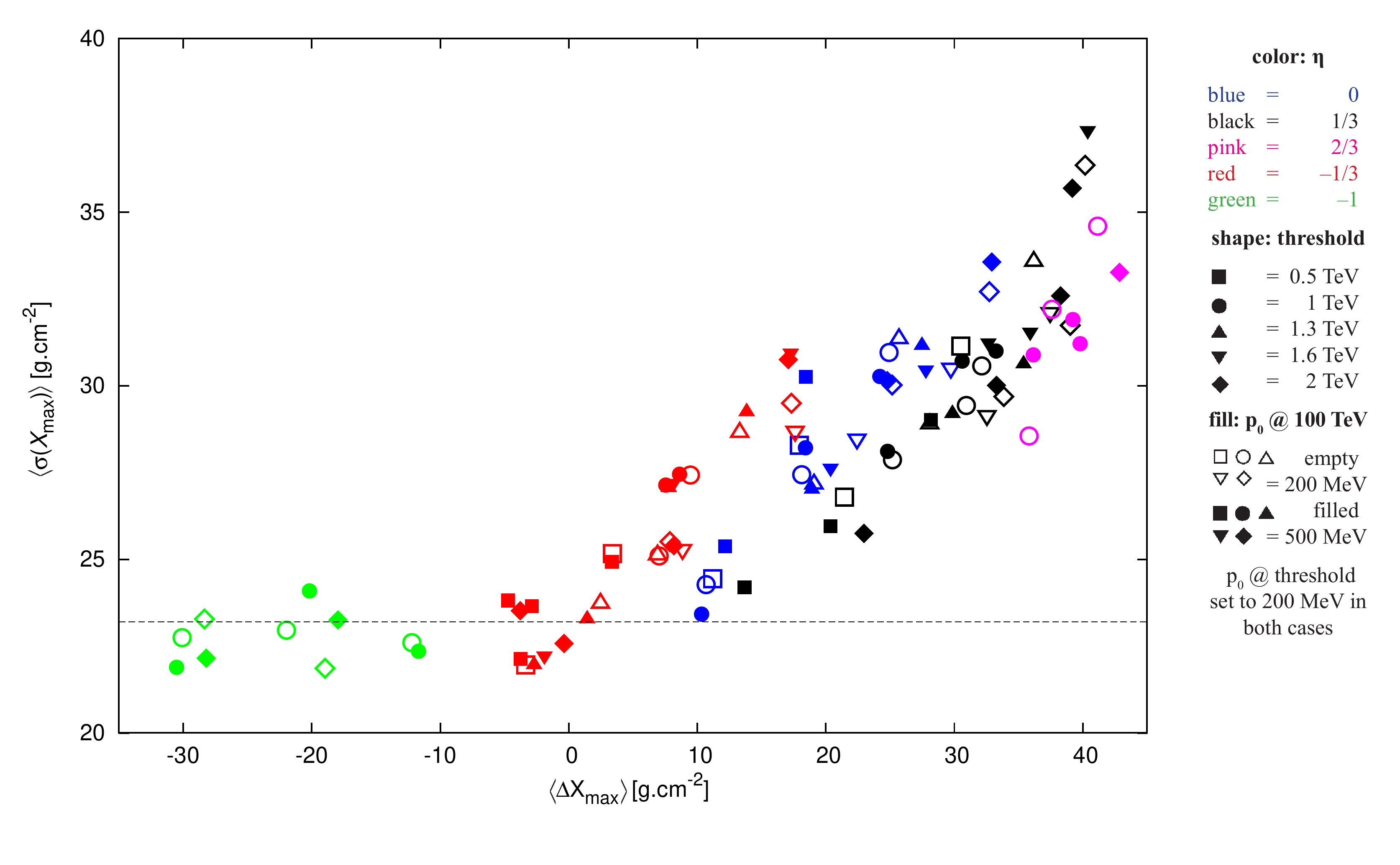}

\caption{$\langle\mathrm{\sigma}(X_\mathrm{max})\rangle$ as a function of $\langle\Delta X_\mathrm{max}\rangle$ for iron showers at $3.2\times 10^{18}$ eV. The plot includes all available simulations at once. The horizontal line corresponds to $\langle\mathrm{\sigma}(X_\mathrm{max})\rangle$ in unmodified iron QGSJETII-4 simulations. For more details on the plots see Sec.~\ref{plots} }
\label{rmsFe}
\end{figure}

For the sake of completeness, we have studied several further variables. Fig.~\ref{rms} shows that the effect of the soft-particle addition on the second moment of the $X_\mathrm{max}$ distribution is small for proton showers: within 4 g\,cm$^{-2}$ for most of the considered choices of parameters, mostly consistent with the accuracy of the simulations themselves. The situation is more complicated for iron showers. Fig.~\ref{muxFe} shows the same relation for them as does Fig.~\ref{mux} for proton showers. Note that while the increase in the number of muons is comparable, the shift in $\langle X_\mathrm{max}\rangle$ has opposite sign for most of the simulations. Should the soft-particle addition model reflect the physical reality, it would almost invariably mean that the average depth of maxima for different primaries are closer to each other than predicted by contemporary models and thus the discrimination power of the fluorescence technique in terms of primary mass composition is smaller than currently assumed. Furthermore note that for iron primaries, the choice of $\eta$ has quite understandably a much larger effect than for proton ones -- even though air is comprised of nuclei, they are quite light. Finally we note that the addition of soft particles can change the variance of $X_\mathrm{max}$ for the iron showers significantly for some choices of parameters, as exhibited in Fig.~\ref{rmsFe}. We however note that the results in this plot might have been influenced by the dependence of the reference frame on the number of wounded nucleons, as described in Sec.~\ref{spam}.

\subsection{Discussion}

The addition of soft particles to high-energy interactions in CR showers does indeed increase both the number of observed muon bundles at DELPHI and the number of muons at 1000 meters from the shower core at ultra-high energies relevant to the Pierre Auger Observatory. The average fraction of c.m.s energy that needs to be converted in each interaction above a given energy threshold to soft particles in order to reach the observed values for the DELPHI benchmarks increases with this energy threshold, as well as with the mean momentum of the added particles, while not showing a strong dependence on the $\eta$ factor related to the amount of wounded nucleons. While the relative number of events with observed multiplicities higher than 20 with respect to QGSJET-01 proton simulations $DPH_{20}$ observed by DELPHI can be, for various thresholds, explained using only 3--7 \% of energy from each interactions, the observed value for the similar quantity $DPH_{80}$ can be reached only for the lower considered thresholds and/or higher fractions of energy converted. However, taking into account the error bands it seems there is a limited parameter region where both predictions can be reconciled.  
It is of interest to note, that both benchmark values correlate rather well with the average total number of particles added over the whole shower -- the choice of energy threshold and other parameters only controls where in the showers the addition occurs, but has a relatively minor impact on the result, if the total number of particles added is the same.

A similar trend can be seen when the same parameter choices are confronted with the constraints imposed by the fact that $X_\mathrm{max}$ data from the Pierre Auger Observatory lie already close to the simulations for proton-induced showers and thus further shift of the simulations towards shallower showers is not desirable. For $DPH_{20}$, the target value is typically reached for shifts in $X_\mathrm{max}$ for proton simulations between 20--30 g\,cm$^{-2}$, which is still compatible within uncertainties with Auger measurements, while for $DPH_{80}$, the shift in proton $X_\mathrm{max}$ would need to be higher to reach the observed value. Also for this variable we can see the same trend where the $X_\mathrm{max}$ shift is lower for lower energy threshold for a given value of the DELPHI benchmarks, because for lower thresholds, such a value requires less energy converted in each interaction. For proton simulations at $3.2\times 10^{18}$ eV we again see a correlation of variables independent of the details of the choice of parameters, namely between the increase in the number of muons on the ground at 1000 meters from the shower axis and the shift of the $X_\mathrm{max}$. If we accept the shift of 30 g\,cm$^{-2}$ as maximal, we can obtain roughly 25 \% more muons on the ground for primary protons.

Some interesting trends can be seen for showers at the energy of $3.2\times 10^{18}$ eV initiated by primary iron nuclei. The $\eta$ factor related to the amount of wounded nucleons becomes important for nucleus-nucleus interactions and while most of the choices have the tendency to deepen the average $X_\mathrm{max}$, both signs of the shift are attainable for iron, while always increasing the predicted number of muons at ground. This does not impose additional restrictions for the model, but could have interesting implications at even higher energies, where the composition appears to be heavier.

\section{Conclusions}

We have shown that particles with small momenta in the c.m.s. of the interaction might be related to both the high-energy muons observed in underground detectors and the general excess of muons observed in ultra-high energy cosmic ray showers. The presented model of adding soft particles into high-energy interactions in CR showers is able to improve the agreement between simulations and data for muon-related observables in CR showers for both the DELPHI detector and the Pierre Auger Observatory. The interplay between this data and the $X_\mathrm{max}$ coming also from Auger restricts the acceptable choices of input parameters for the model.

\section*{Acknowledgements}
This work was supported by the grants of the Ministry of Education of the Czech Republic (MSMT-CR LG13007 and LG15014). The authors are thankful to Dalibor Nosek and Petr Tr\'{a}vn\'{i}\v{c}ek for their invaluable advice and suggestions, as well as to members of the Pierre Auger Collaboration for their helpful comments. We appreciate access to computational resources provided by Computing center of the Institute of Physics of CAS \cite{farma} and supported by project LG13007.

\section*{References}
\bibliography{spam}

\begin{thebibliography}{10}
\expandafter\ifx\csname url\endcsname\relax
  \def\url#1{\texttt{#1}}\fi
\expandafter\ifx\csname urlprefix\endcsname\relax\def\urlprefix{URL }\fi
\expandafter\ifx\csname href\endcsname\relax
  \def\href#1#2{#2} \def\path#1{#1}\fi

\bibitem{QGSII-4}
S.~Ostapchenko,
  \href{http://link.aps.org/doi/10.1103/PhysRevD.83.014018}{{Monte Carlo}
  treatment of hadronic interactions in enhanced {Pomeron} scheme: {QGSJET-II}
  model}, Phys. Rev. D 83 (2011) 014018.
\newblock \href {http://dx.doi.org/10.1103/PhysRevD.83.014018}
  {\path{doi:10.1103/PhysRevD.83.014018}}.
\newline\urlprefix\url{http://link.aps.org/doi/10.1103/PhysRevD.83.014018}

\bibitem{EPOS-LHC}
T.~Pierog, I.~Karpenko, J.~M. Katzy, E.~Yatsenko, K.~Werner,
  \href{http://link.aps.org/doi/10.1103/PhysRevC.92.034906}{{EPOS LHC}: Test of
  collective hadronization with data measured at the {CERN} {Large Hadron
  Collider}}, Phys. Rev. C 92 (2015) 034906.
\newblock \href {http://dx.doi.org/10.1103/PhysRevC.92.034906}
  {\path{doi:10.1103/PhysRevC.92.034906}}.
\newline\urlprefix\url{http://link.aps.org/doi/10.1103/PhysRevC.92.034906}

\bibitem{Sibyll}
E.-J. Ahn, R.~Engel, T.~K. Gaisser, P.~Lipari, T.~Stanev,
  \href{http://link.aps.org/doi/10.1103/PhysRevD.80.094003}{Cosmic ray
  interaction event generator {SIBYLL} 2.1}, Phys. Rev. D 80 (2009) 094003.
\newblock \href {http://dx.doi.org/10.1103/PhysRevD.80.094003}
  {\path{doi:10.1103/PhysRevD.80.094003}}.
\newline\urlprefix\url{http://link.aps.org/doi/10.1103/PhysRevD.80.094003}

\bibitem{L3C2}
C.~Timmermans,
  \href{http://www.sciencedirect.com/science/article/pii/B9780444513434500326}{Cosmic
  rays with the {LEP} detectors}, in: S.~Laenen, P.~Bentvelsen, J.~de~Jong,
  E.~Koch (Eds.), Proceedings of the 31st International Conference on High
  Energy Physics Ichep 2002, JAI, Amsterdam, 2003, pp. 109 -- 112.
\newblock \href
  {http://dx.doi.org/http://dx.doi.org/10.1016/B978-0-444-51343-4.50032-6}
  {\path{doi:http://dx.doi.org/10.1016/B978-0-444-51343-4.50032-6}}.
\newline\urlprefix\url{http://www.sciencedirect.com/science/article/pii/B9780444513434500326}

\bibitem{ALEPH}
V.~Avati, L.~Dick, K.~Eggert, J.~Strom, H.~Wachsmuth, et~al., {Cosmic
  multi-muon events observed in the underground {CERN-LEP} tunnel with the
  {ALEPH} experiment}, Astropart.Phys. 19 (2003) 513--523.
\newblock \href {http://dx.doi.org/10.1016/S0927-6505(02)00247-5}
  {\path{doi:10.1016/S0927-6505(02)00247-5}}.

\bibitem{DELPHI}
J.~Abdallah, et~al., {Study of multi-muon bundles in cosmic ray showers
  detected with the {DELPHI} detector at {LEP}}, Astropart.Phys. 28 (2007)
  273--286.
\newblock \href {http://arxiv.org/abs/0706.2561} {\path{arXiv:0706.2561}},
  \href {http://dx.doi.org/10.1016/j.astropartphys.2007.06.001}
  {\path{doi:10.1016/j.astropartphys.2007.06.001}}.

\bibitem{ALICE2}
J.~Adam, et~al., {Study of cosmic ray events with high muon multiplicity using
  the {ALICE} detector at the {CERN} {Large Hadron Collider}}, JCAP 1601~(01)
  (2016) 032.
\newblock \href {http://arxiv.org/abs/1507.07577} {\path{arXiv:1507.07577}},
  \href {http://dx.doi.org/10.1088/1475-7516/2016/01/032}
  {\path{doi:10.1088/1475-7516/2016/01/032}}.

\bibitem{NEVOD}
A.~G. Bogdanov, et~al., {Investigation of the energy characteristics of {EAS}
  muon component with the {NEVOD-DECOR} setup}, J. Phys. Conf. Ser. 675~(3)
  (2016) 032035.
\newblock \href {http://dx.doi.org/10.1088/1742-6596/675/3/032035}
  {\path{doi:10.1088/1742-6596/675/3/032035}}.

\bibitem{QGSII-3}
S.~Ostapchenko, Hadronic interactions at cosmic ray energies, Nuclear Physics B
  -- Proceedings Supplements 175–176~(0) (2008) 73--80, proceedings of the XIV
  International Symposium on Very High Energy Cosmic Ray Interactions.
\newblock \href {http://dx.doi.org/10.1016/j.nuclphysbps.2007.10.011}
  {\path{doi:10.1016/j.nuclphysbps.2007.10.011}}.

\bibitem{AUGER}
{The Pierre Auger Collaboration},
  \href{http://www.sciencedirect.com/science/article/pii/S0168900215008086}{The
  {Pierre Auger Cosmic Ray Observatory}}, Nuclear Instruments and Methods in
  Physics Research Section A: Accelerators, Spectrometers, Detectors and
  Associated Equipment 798 (2015) 172 -- 213.
\newblock \href
  {http://dx.doi.org/http://dx.doi.org/10.1016/j.nima.2015.06.058}
  {\path{doi:http://dx.doi.org/10.1016/j.nima.2015.06.058}}.
\newline\urlprefix\url{http://www.sciencedirect.com/science/article/pii/S0168900215008086}

\bibitem{AUGERmu1}
{The Pierre Auger Collaboration},
  \href{http://link.aps.org/doi/10.1103/PhysRevD.91.032003}{Muons in air
  showers at the {Pierre Auger Observatory}: Mean number in highly inclined
  events}, Phys. Rev. D 91 (2015) 032003.
\newblock \href {http://dx.doi.org/10.1103/PhysRevD.91.032003}
  {\path{doi:10.1103/PhysRevD.91.032003}}.
\newline\urlprefix\url{http://link.aps.org/doi/10.1103/PhysRevD.91.032003}

\bibitem{AUGERmu2}
{G. Farrar for the Pierre Auger Collaboration}, The muon content of hybrid
  events recorded at the {Pierre Auger Observatory}, Proceedings of the 33rd
  ICRC\href {http://arxiv.org/abs/1307.5059} {\path{arXiv:1307.5059}}.

\bibitem{AUGERmu3}
{B. Kegl for the Pierre Auger Collaboration}, Measurement of the muon signal
  using the temporal and spectral structure of the signals in surface detectors
  of the {Pierre Auger Observatory}, Proceedings of the 33rd ICRC\href
  {http://arxiv.org/abs/1307.5059} {\path{arXiv:1307.5059}}.

\bibitem{Nosek1}
J.~\v{R}\'{i}dk\'{y}, P.~Tr\'{a}vn\'{i}\v{c}ek, P.~Ne\v{c}esal, D.~Nosek,
  \href{http://inspirehep.net/record/753135/files/arXiv:0706.2145.pdf}{{Prompt
  muons in extended air showers}}, in: {Proceedings, 30th International Cosmic
  Ray Conference (ICRC 2007)}, Vol.~4, 2007, pp. 605--608, [4,605(2007)].
\newblock \href {http://arxiv.org/abs/0706.2145} {\path{arXiv:0706.2145}}.
\newline\urlprefix\url{http://inspirehep.net/record/753135/files/arXiv:0706.2145.pdf}

\bibitem{Charm}
A.~R. et~al. (The LHCb~Collaboration),
  \href{http://www.sciencedirect.com/science/article/pii/S0550321313000965}{Prompt
  charm production in pp collisions at sqrt(s)=7 {TeV}}, Nuclear Physics B
  871~(1) (2013) 1 -- 20.
\newblock \href
  {http://dx.doi.org/http://dx.doi.org/10.1016/j.nuclphysb.2013.02.010}
  {\path{doi:http://dx.doi.org/10.1016/j.nuclphysb.2013.02.010}}.
\newline\urlprefix\url{http://www.sciencedirect.com/science/article/pii/S0550321313000965}

\bibitem{Necesal}
P.~Ne\v{c}esal, PhD Thesis, 2014, institute of Physics, Czech Academy of
  Sciences.

\bibitem{DELSIM}
D.~Collaboration, {DELSIM, Delphi Event Generation and Detector Simulation
  User's guide}, DELPHI NOTE: 89-67 PROG 142.

\bibitem{Travnicek}
P.~Tr\'{a}vn\'{i}\v{c}ek, PhD Thesis, 2004, institute of Physics, Czech Academy
  of Sciences.

\bibitem{QGS01a}
N.~Kalmykov, S.~Ostapchenko, A.~Pavlov,
  \href{http://www.sciencedirect.com/science/article/pii/S0920563296008468}{Quark-gluon-string
  model and {EAS} simulation problems at ultra-high energies}, Nuclear Physics
  B - Proceedings Supplements 52~(3) (1997) 17 -- 28.
\newblock \href
  {http://dx.doi.org/http://dx.doi.org/10.1016/S0920-5632(96)00846-8}
  {\path{doi:http://dx.doi.org/10.1016/S0920-5632(96)00846-8}}.
\newline\urlprefix\url{http://www.sciencedirect.com/science/article/pii/S0920563296008468}

\bibitem{QGS01b}
N.~N. Kalmykov, S.~S. Ostapchenko, {The Nucleus-nucleus interaction, nuclear
  fragmentation, and fluctuations of extensive air showers}, Phys. Atom. Nucl.
  56 (1993) 346--353, [Yad. Fiz.56N3,105(1993)].

\bibitem{Corsika}
D.~{Heck}, et~al., FZKA Report Forschungszentrum Karlsruhe 6019.

\bibitem{KASCADE}
A.~K.-G.~C. Haungs, High-energy cosmic rays measured with {KASCADE-Grande},
  Proceedings of the 33rd ICRC\href {http://arxiv.org/abs/1308.1485}
  {\path{arXiv:1308.1485}}.

\bibitem{geisha}
H.~{Fesefeldt}, Report RWTH Aachen PITHA-85/02.

\bibitem{L3C}
Recent results from {L3+COSMICS} at {CERN} {L3} collaboration, Nuclear Physics
  B - Proceedings Supplements 110~(0) (2002) 469--471.
\newblock \href {http://dx.doi.org/10.1016/S0920-5632(02)01537-2}
  {\path{doi:10.1016/S0920-5632(02)01537-2}}.

\bibitem{LHCf}
T.~L. Collaboration, \href{http://stacks.iop.org/1748-0221/3/i=08/a=S08006}{The
  {LHCf} detector at the {CERN Large Hadron Collider}}, Journal of
  Instrumentation 3~(08) (2008) S08006.
\newline\urlprefix\url{http://stacks.iop.org/1748-0221/3/i=08/a=S08006}

\bibitem{QGP}
J.~Ridky,
  \href{http://www.sciencedirect.com/science/article/pii/S092765050100161X}{Can
  we observe the quark gluon plasma in cosmic ray showers}, Astroparticle
  Physics 17~(3) (2002) 355 -- 365.
\newblock \href
  {http://dx.doi.org/http://dx.doi.org/10.1016/S0927-6505(01)00161-X}
  {\path{doi:http://dx.doi.org/10.1016/S0927-6505(01)00161-X}}.
\newline\urlprefix\url{http://www.sciencedirect.com/science/article/pii/S092765050100161X}

\bibitem{Nosek2}
J.~\v{R}\'{i}dk\'{y}, D.~Nosek,
  \href{http://www.sciencedirect.com/science/article/pii/S0920563204006322}{On
  sensitivity of {Cherenkov} radiation to the dynamics of high-energy cosmic
  ray interactions}, Nuclear Physics B - Proceedings Supplements 138 (2005) 299
  -- 302, proceedings of the Eighth International Workshop on Topics in
  Astroparticle and Undeground Physics.
\newblock \href
  {http://dx.doi.org/http://dx.doi.org/10.1016/j.nuclphysbps.2004.11.067}
  {\path{doi:http://dx.doi.org/10.1016/j.nuclphysbps.2004.11.067}}.
\newline\urlprefix\url{http://www.sciencedirect.com/science/article/pii/S0920563204006322}

\bibitem{Xmax}
{The Pierre Auger Collaboration},
  \href{http://link.aps.org/doi/10.1103/PhysRevD.90.122005}{Depth of maximum of
  air-shower profiles at the {Pierre Auger Observatory}. {I}. measurements at
  energies above $1{0}^{17.8}\text{ }\mathrm{eV}$}, Phys. Rev. D 90 (2014)
  122005.
\newblock \href {http://dx.doi.org/10.1103/PhysRevD.90.122005}
  {\path{doi:10.1103/PhysRevD.90.122005}}.
\newline\urlprefix\url{http://link.aps.org/doi/10.1103/PhysRevD.90.122005}

\bibitem{ALICE}
A.~Collaboration,
  \href{http://www.sciencedirect.com/science/article/pii/S0370269310009731}{Transverse
  momentum spectra of charged particles in proton–proton collisions at with
  {ALICE} at the {LHC}}, Physics Letters B 693~(2) (2010) 53 -- 68.
\newblock \href
  {http://dx.doi.org/http://dx.doi.org/10.1016/j.physletb.2010.08.026}
  {\path{doi:http://dx.doi.org/10.1016/j.physletb.2010.08.026}}.
\newline\urlprefix\url{http://www.sciencedirect.com/science/article/pii/S0370269310009731}

\bibitem{TOTEM}
G.~Antchev, et~al., {Measurement of the forward charged particle pseudorapidity
  density in pp collisions at $\sqrt{s} = 8$ TeV using a displaced interaction
  point}, Eur. Phys. J. C75~(3) (2015) 126.
\newblock \href {http://arxiv.org/abs/1411.4963} {\path{arXiv:1411.4963}},
  \href {http://dx.doi.org/10.1140/epjc/s10052-015-3343-7}
  {\path{doi:10.1140/epjc/s10052-015-3343-7}}.

\bibitem{Ridky}
J.~Ridky, \href{http://dx.doi.org/10.1002/prop.2190361002}{Is there a hierarchy
  of the hadron low $p_{T}$ spectra?}, Fortschritte der Physik/Progress of
  Physics 36~(10) (1988) 707--780.
\newblock \href {http://dx.doi.org/10.1002/prop.2190361002}
  {\path{doi:10.1002/prop.2190361002}}.
\newline\urlprefix\url{http://dx.doi.org/10.1002/prop.2190361002}

\bibitem{Gaisser-Hillas}
H.~A. Gaisser~T.K., in: {Proceedings, 15th International Cosmic Ray
  Conference}, 1977, p. 358, plovdiv, Bulgaria.

\bibitem{NK}
N.~J. Kamata~K., Prog. Theoret. Phys. Suppl. 6 (1958) 93.

\bibitem{G}
G.~K., Wilson J. G. (ed), Progress in cosmic ray physics III (1956) 3.

\bibitem{Ulrich}
R.~Ulrich, R.~Engel, M.~Unger,
  \href{http://link.aps.org/doi/10.1103/PhysRevD.83.054026}{Hadronic
  multiparticle production at ultrahigh energies and extensive air showers},
  Phys. Rev. D 83 (2011) 054026.
\newblock \href {http://dx.doi.org/10.1103/PhysRevD.83.054026}
  {\path{doi:10.1103/PhysRevD.83.054026}}.
\newline\urlprefix\url{http://link.aps.org/doi/10.1103/PhysRevD.83.054026}

\bibitem{farma}
D.~Adamova, J.~Chudoba, M.~Elias, L.~Fiala, T.~Kouba, M.~Lokajicek, J.~Svec,
  \href{http://stacks.iop.org/1742-6596/608/i=1/a=012035}{{WLCG} {Tier-2} site
  in {Prague}: a little bit of history, current status and future
  perspectives}, Journal of Physics: Conference Series 608~(1) (2015) 012035.
\newline\urlprefix\url{http://stacks.iop.org/1742-6596/608/i=1/a=012035}

\end{thebibliography}

\end{document}